\documentclass[%
 aip,
 amsmath,amssymb,
 reprint,%
]{revtex4-1}
\usepackage{graphicx}% Include figure files
\usepackage{dcolumn}% Align table columns on decimal point
\usepackage{bm}% bold math
\usepackage[utf8]{inputenc}
\usepackage[T1]{fontenc}
\usepackage{mathptmx}
\usepackage{color}
\usepackage[normalem]{ulem}

\begin{document}

\preprint{PIPER}

\title[PIPER receiver]{Sub-Kelvin cooling for two kilopixel bolometer arrays in the PIPER receiver}
% Force line breaks with \\

\author{E. R. Switzer}
 \email{eric.r.switzer@nasa.gov}
 \affiliation{NASA Goddard Space Flight Center, Greenbelt, MD 20771, USA}

\author{P. A. R. Ade}
\affiliation{School of Physics and Astronomy, Cardiff University, Queens Buildings, The Parade, Cardiff, CF24 3AA, UK}

\author{T. Baildon}
\affiliation{Department of Physics, University of Michigan, Ann Arbor, MI, 48109, USA}

\author{D. Benford}
\affiliation{NASA Headquarters, Washington, DC 20546, USA}

\author{C. L. Bennett}
\affiliation{Department of Physics and Astronomy, Johns Hopkins University, Baltimore, MD 21218, USA}

\author{D. T. Chuss}
\affiliation{Department of Physics, Villanova University, Villanova, PA 19085, USA}

\author{R. Datta}
\affiliation{Department of Physics and Astronomy, Johns Hopkins University, Baltimore, MD 21218, USA}

%\author{J. L. Dotson}
%\affiliation{NASA Ames Research Center, Moffett Blvd, Mountain View, CA 94035}

\author{J. R. Eimer}
\affiliation{Department of Physics and Astronomy, Johns Hopkins University, Baltimore, MD 21218, USA}

\author{D. J. Fixsen}
\affiliation{University of Maryland, College Park, MD 20742, USA}
\affiliation{NASA Goddard Space Flight Center, Greenbelt, MD 20771, USA}

\author{N. N. Gandilo}
\altaffiliation[]{Current address: Steward Observatory, University of Arizona, Tucson, AZ 85721, USA}
\affiliation{Department of Physics and Astronomy, Johns Hopkins University, Baltimore, MD 21218, USA}

\author{T. M. Essinger-Hileman}
\affiliation{NASA Goddard Space Flight Center, Greenbelt, MD 20771, USA}

\author{M. Halpern}
\affiliation{Department of Physics and Astronomy, University of British Columbia, Vancouver, BC V6T 1Z4, Canada}

\author{G. Hilton}
\affiliation{National Institute of Standards and Technology, Boulder, CO 80305, USA}

%\author{G. F. Hinshaw}
%\affiliation{Department of Physics and Astronomy, University of British Columbia, Vancouver, BC V6T 1Z4, Canada}

\author{K. Irwin}
\affiliation{Kavli Institute for Particle Astrophysics and Cosmology, Stanford University, Stanford, CA 94305}
\affiliation{Department of Physics, Stanford University, Stanford, CA 94305}
\affiliation{SLAC National Accelerator Laboratory, Menlo Park, CA 94025}

\author{C. Jhabvala}
\affiliation{NASA Goddard Space Flight Center, Greenbelt, MD 20771, USA}

\author{M. Kimball}
\affiliation{NASA Goddard Space Flight Center, Greenbelt, MD 20771, USA}

\author{A. Kogut}
\affiliation{NASA Goddard Space Flight Center, Greenbelt, MD 20771, USA}

\author{J. Lazear}
\affiliation{Department of Physics and Astronomy, Johns Hopkins University, Baltimore, MD 21218, USA}
%\affiliation{Northrop Grumman, Redondo Beach, CA 90278}

\author{L. N. Lowe}
\affiliation{Sigma Space, a subsidiary of Hexagon US Federal, Lanham, MD 20706}
\affiliation{NASA Goddard Space Flight Center, Greenbelt, MD 20771, USA}

\author{J. J. McMahon}
\affiliation{Department of Physics, University of Michigan, Ann Arbor, MI, 48109, USA}

\author{T. M. Miller}
\affiliation{NASA Goddard Space Flight Center, Greenbelt, MD 20771, USA}

\author{P. Mirel}
\affiliation{Sigma Space, a subsidiary of Hexagon US Federal, Lanham, MD 20706}
\affiliation{NASA Goddard Space Flight Center, Greenbelt, MD 20771, USA}
% \altaffiliation[]{Sigma Space, a subsidiary of Hexagon US Federal, Lanham, MD 20706}

\author{S. H. Moseley}
\affiliation{NASA Goddard Space Flight Center, Greenbelt, MD 20771, USA}

\author{S. Pawlyk}
\affiliation{University of Maryland, College Park, MD 20742, USA}
\affiliation{NASA Goddard Space Flight Center, Greenbelt, MD 20771, USA}

\author{S. Rodriguez}
\affiliation{NASA Goddard Space Flight Center, Greenbelt, MD 20771, USA}

\author{E. Sharp}
\affiliation{GST Inc., Greenbelt, MD 20770}
\affiliation{NASA Goddard Space Flight Center, Greenbelt, MD 20771, USA}

\author{P. Shirron}
\affiliation{NASA Goddard Space Flight Center, Greenbelt, MD 20771, USA}

\author{J. G. Staguhn}
\affiliation{NASA Goddard Space Flight Center, Greenbelt, MD 20771, USA}
\affiliation{Department of Physics and Astronomy, Johns Hopkins University, Baltimore, MD 21218, USA}

\author{D. F. Sullivan}
\affiliation{NASA Goddard Space Flight Center, Greenbelt, MD 20771, USA}

% ATA Aerospace. Their address is 7474 Greenway Center Drive Suite 500, Greenbelt, MD 20770 301-329-8200
\author{P. Taraschi}
\affiliation{ATA Aerospace, Greenbelt, MD 20770}
%\altaffiliation[]{Sigma Space, a subsidiary of Hexagon US Federal, Lanham, MD 20706}
\affiliation{NASA Goddard Space Flight Center, Greenbelt, MD 20771, USA}

\author{C. E. Tucker}
\affiliation{School of Physics and Astronomy, Cardiff University, Queens Buildings, The Parade, Cardiff, CF24 3AA, UK}

\author{A. Walts}
\affiliation{NASA Goddard Space Flight Center, Greenbelt, MD 20771, USA}

\author{E. J. Wollack}
\affiliation{NASA Goddard Space Flight Center, Greenbelt, MD 20771, USA}

\date{\today}% It is always \today, today,

\begin{abstract}
The Primordial Inflation Polarization Explorer (PIPER) is a balloon-borne telescope mission to search for inflationary gravitational waves from the early universe. PIPER employs two $32 \times 40$ arrays of superconducting transition-edge sensors, which operate at $100$\,mK. An open bucket dewar of liquid helium maintains the receiver and telescope optics at $1.7$\,K. We describe the thermal design of the receiver and sub-kelvin cooling with a continuous adiabatic demagnetization refrigerator (CADR). The CADR operates between $70-130$\,mK and provides ${\approx}10\,\mu {\rm W}$ cooling power at $100$\,mK, nearly five times the loading of the two detector assemblies. We describe electronics and software to robustly control the CADR, overall CADR performance in flight-like integrated receiver testing, and practical considerations for implementation in the balloon float environment.
\end{abstract}
% and results from integrated receiver testing and an October 2017 engineering flight.

\maketitle

\section{\label{sec:intro}Introduction}

Large cryogenic detector arrays provide revolutionary sensitivity in modern instruments for surveys across the electromagnetic spectrum, from centimeter wavelengths to calorimetry of energetic particles. Transition edge sensor (TES) and kinetic inductance detector (KIDs) detection approaches in these instruments require sub-Kelvin operating temperatures to reach required sensitivity. For temperatures below approximately $300$\,mK, adiabatic demagnetization refrigeration (ADR) and dilution refrigeration are the leading approaches and have been used in space applications, with recent examples in Hitomi\cite{2018JATIS...4b1403S} and Planck\cite{2011A&A...536A...2P}. The cooling mechanism of the ADR is solid-state and has no moving parts, making it mechanically robust and long-lived for space applications. A continuous ADR (CADR) is a variant that further provides isothermal cooling power that remains uninterrupted through the ADR cycle.

Here we describe the receiver and sub-Kelvin cooling for the Primordial Inflation Polarization Explorer (PIPER)\cite{2010SPIE.7733E..3BE, 2012SPIE.8452E..1JK, 2014RScI...85f4501C, 2018SPIE10708E..06P}. PIPER's mission is to map the polarization of the cosmic microwave background (CMB) in a search for inflationary B-modes on large angular scales. During day-long conventional balloon flights, it will map the CMB and Milky Way polarized dust emission in four frequency channels from $200-600$\,GHz. 

PIPER consists of twin co-pointed telescopes cooled to $1.5-1.7$\,K within a large liquid helium (LHe) bucket dewar. The LHe is pumped by the atmosphere at the balloon float altitude of $30-37$\,km. A variable-delay polarization modulator (VPM) on each telescope modulates between linear and circular polarization through the translation of a mirror behind a wire grid\citep{2014RScI...85f4501C}. Three advantages of this cryogenic modulation scheme are that 1) the wire grid is large enough to be the first optical element at the 29\,cm entrance pupil (with 40\,cm open aperture)\citep{2010SPIE.7733E..3BE}, 2) small linear (rather than rotary) translations allow $4$\,Hz modulation and 3) a cryogenic wire grid will not emit in-band. Silicon reimaging optics\cite{2013ApOpt..52.8747D} feed two $32 \times 40$ TES focal planes\cite{2014SPIE.9153E..3CJ}. For sub-Kelvin cooling, PIPER uses a CADR to maintain the detector arrays at $70-130$\,mK.

The PIPER receiver presents unique challenges which are not present in traditional cryogenic designs.  Sec.\,\ref{sec:overview} reviews the architecture and constraints of the receiver. Sec.\,\ref{sec:subkcooling} describes sub-Kelvin cooling in the receiver, followed by a description of the CADR drive electronics, harnesses, and control software in Sec.\,\ref{sec:cadrelect}. Sec.\,\ref{sec:cadroperation} outlines CADR control and operational considerations and Sec.\,\ref{sec:performance} reviews the cryogenic performance of the CADR in combination with the TES arrays. 

This publication is the first description of CADR performance in conjunction with large TES arrays for continuum radiation. The previous literature\citep{2009AIPC.1185..442I} describes preliminary CADR operation for TES microcalorimeters. The MAX/MAXIMA/MAXIPOL \cite{1993SPIE.1946..110T, 1994InPhT..35..431R, 2006RScI...77g1101R, 2007ApJ...665...42J} and MSAM2 \citep{MSAM2, 1995ApL&C..32..273K} balloon missions for the CMB employed single-shot ADRs based on ferric ammonium alum salts, which derive from prototypes for the Space Infrared Telescope Facility \citep{1995Cryo...35..303H, 1990Cryo...30..271T}. 
Sounding rockets for X-ray calorimetry \citep{1996NIMPA.370..266M, 2016JLTP..184..699G} have also employed single-shot ADRs.

Results on CADR performance are from flight-like tests of the integrated receiver at NASA Goddard Space Flight Center in June 2018. The 2018 Ft. Sumner, New Mexico flight campaign was terminated before a PIPER launch opportunity, and the integrated instrument is expected to fly in Fall 2019 from Ft. Sumner. Data from an October 2017 engineering flight from Ft. Sumner demonstrate the cryogenic environment at float altitude, operation of the flight electronics, and recovery after landing. That flight fielded one engineering detector array and demonstrated the CADR in single-shot mode with flight electronics. A cold-open contact that appeared after shipping (Sec.\,\ref{sec:cadroperation}) limited demonstration of continuous operation in flight.

\section{Receiver Overview\label{sec:overview}}

\subsection{Overview and operational environment \label{ssec:flightenv}}

\begin{figure}
\includegraphics[scale=0.22]{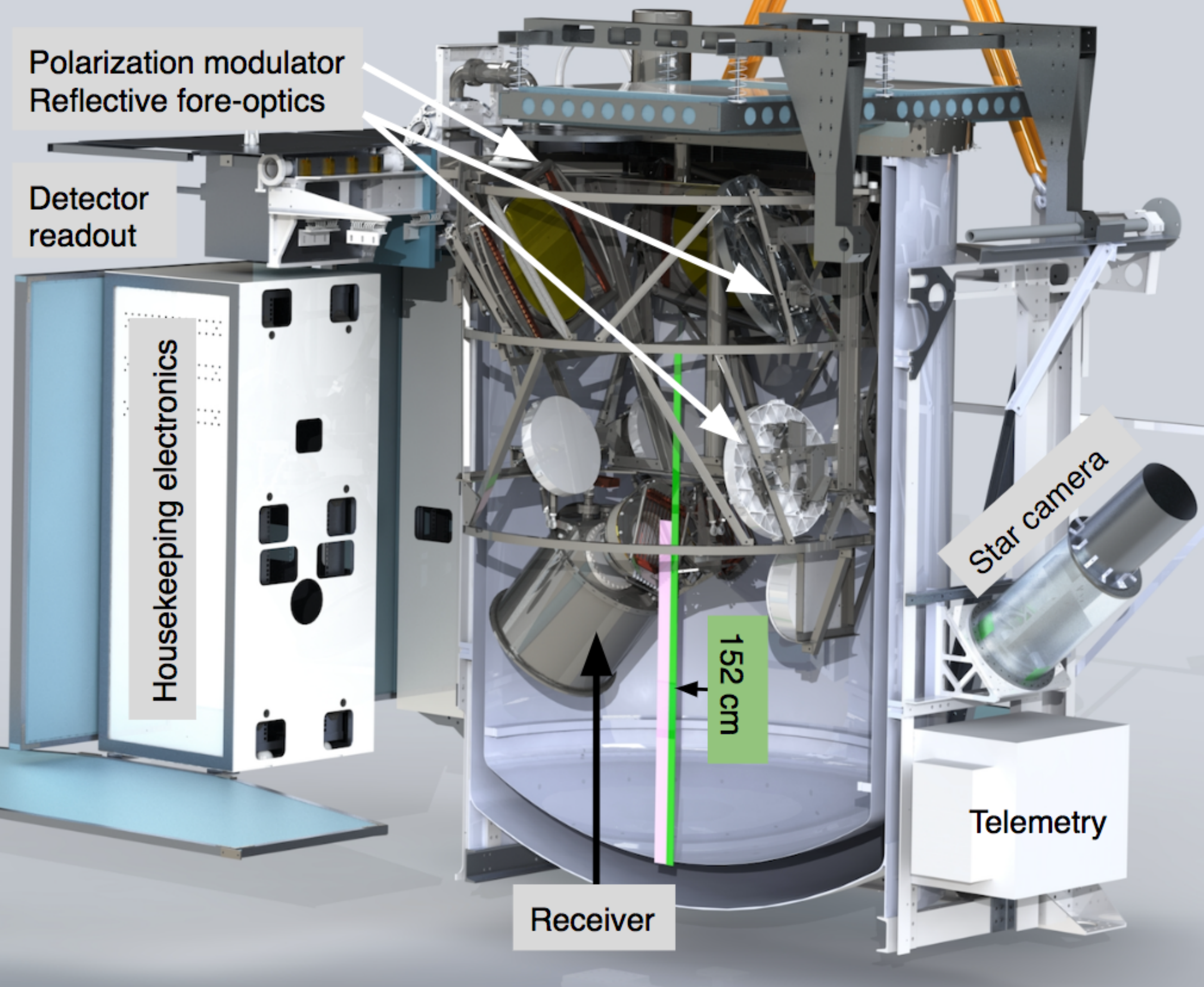}
\caption{\label{fig:piper_section} PIPER is a balloon-borne polarimeter designed to search for inflationary gravitational waves in observations from $200-600$\,GHz. The payload is $2360$\,kg (dry) and is built around a dewar filled with 3000\,L LHe. A variable-delay polarization modulator (VPM) precedes the off-axis telescope and silicon reimaging optics that feed two $32\times 40$ TES arrays. The twin sides of the telescope measure Stokes Q and U, respectively. The pink and green lines indicate LHe level gauges, which are used to show scale.}
\end{figure}

A large, open bucket dewar (Fig.\,\ref{fig:piper_section}) houses the twin telescopes and ascends to $36$\,km target altitude on a 34 million cubic foot (heavy variant) balloon. PIPER launches with ${\approx}3000$\,L of LHe, and the dewar is $1.5$\,m inner diameter by $2$\,m deep. A hatch covers two optical exit ports on the ground and opens at float altitude.

\begin{figure}
\includegraphics[scale=0.50]{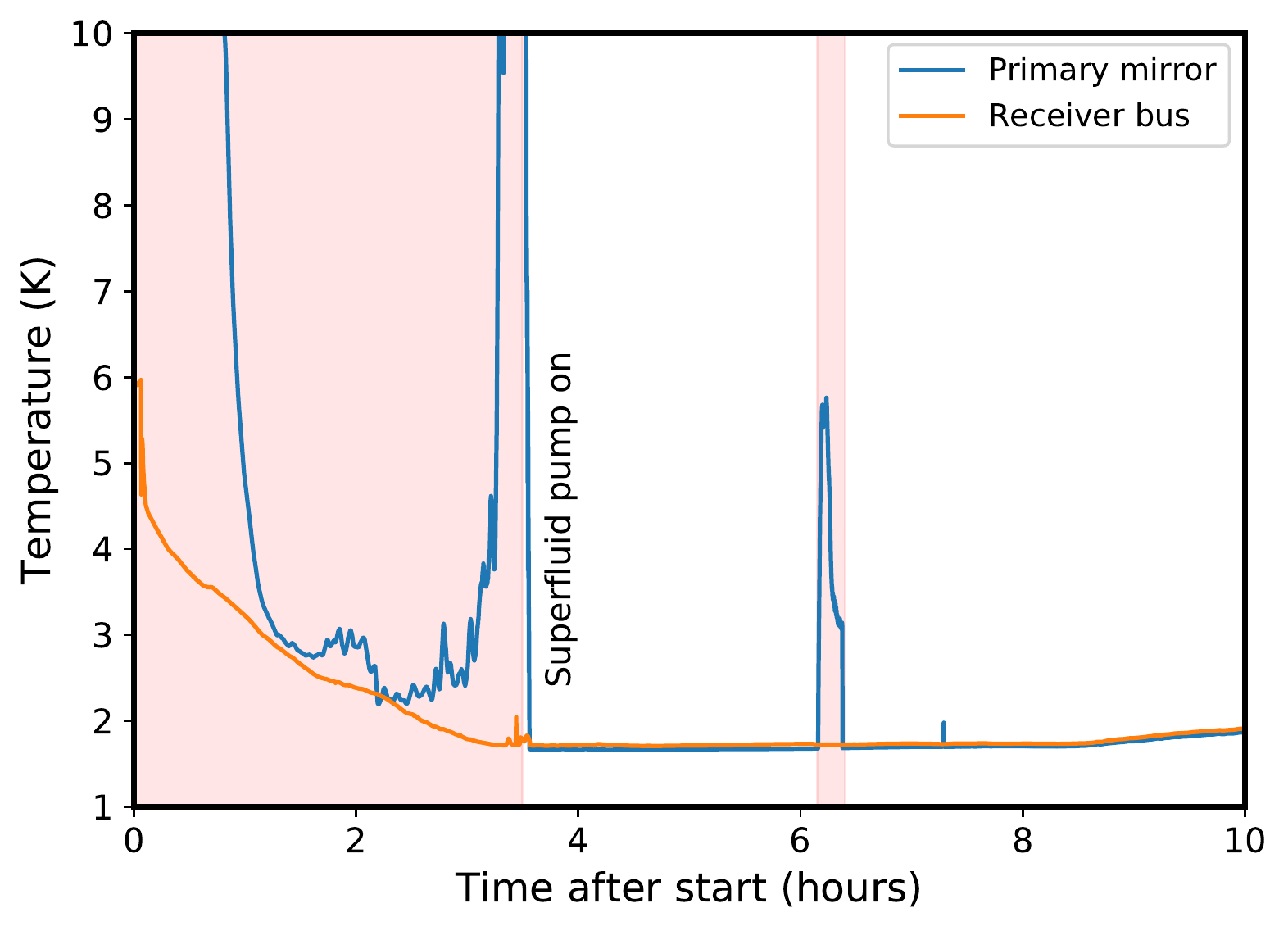}
\caption{\label{fig:piper_ascent} Temperature data from the PIPER 2017 engineering flight. During the ascent, rapid helium gas evolution carries heat away from the receiver and telescope. LHe becomes a superfluid at altitudes above approximately $24$\,km, which allows operation of the superfluid pumps. Red regions indicate times when the superfluid pump was not operating. Ascent to float altitude requires approximately 3\,hr. After the ascent, the superfluid bath stabilizes, and the receiver and optics remain in a weak vacuum above the superfluid. Without superfluid pumps, the telescope temperature rises quickly from parasitic loading, with the primary approaching $35$\,K. (The optics remain sufficiently cold to contribute negligibly to the photon background if the pumps fail.) Both optics and receiver remain at $1.7$\,K when the pumps are on. Approximately 6 hours into the engineering flight, the pumps were turned off to assess their impact.}
\end{figure}

Using heritage from the ARCADE mission\cite{2004ApJS..154..493K} 1) superfluid pumps cool the optics to ${\approx}2$\,K, and 2) high gas flow keeps the optics clean without the need for an emissive window. The helium loss rate at float was measured to be $105$\,L/hr in the 2017 engineering flight, corresponding to $7\,{\rm m}^3/{\rm s}$ gas evolution. Boil-off during ascent consumes approximately $30\%$ of the initial helium fill, so an initial fill of $3000$\,L results in a 20\,hr hold at balloon float altitude. This duration is sufficient for the approximately $12$\,hr science period from sunset to sunrise in a conventional campaign. The passive loss rate while waiting for the flight is $220$\,L/day. Fig.\,\ref{fig:piper_ascent} shows the receiver and optics temperatures during the ascent of the engineering flight. 

% $30$\,km float?

All structural elements in the dewar are stainless steel, which provides low heat conduction through the walls of the dewar and frame of the telescope, as well as homogeneous contraction through a shared coefficient of thermal expansion (CTE). Only the mirrors are aluminum, and mount flexures maintain alignment while taking up the differential CTE with the frame. The receiver houses two $32 \times 40$ TES focal planes that are cooled by the CADR to a nominal bath temperature of $100$\,mK. 

\subsection{Receiver Architecture \label{ssec:submarine}}

\begin{figure}
\includegraphics[scale=0.22]{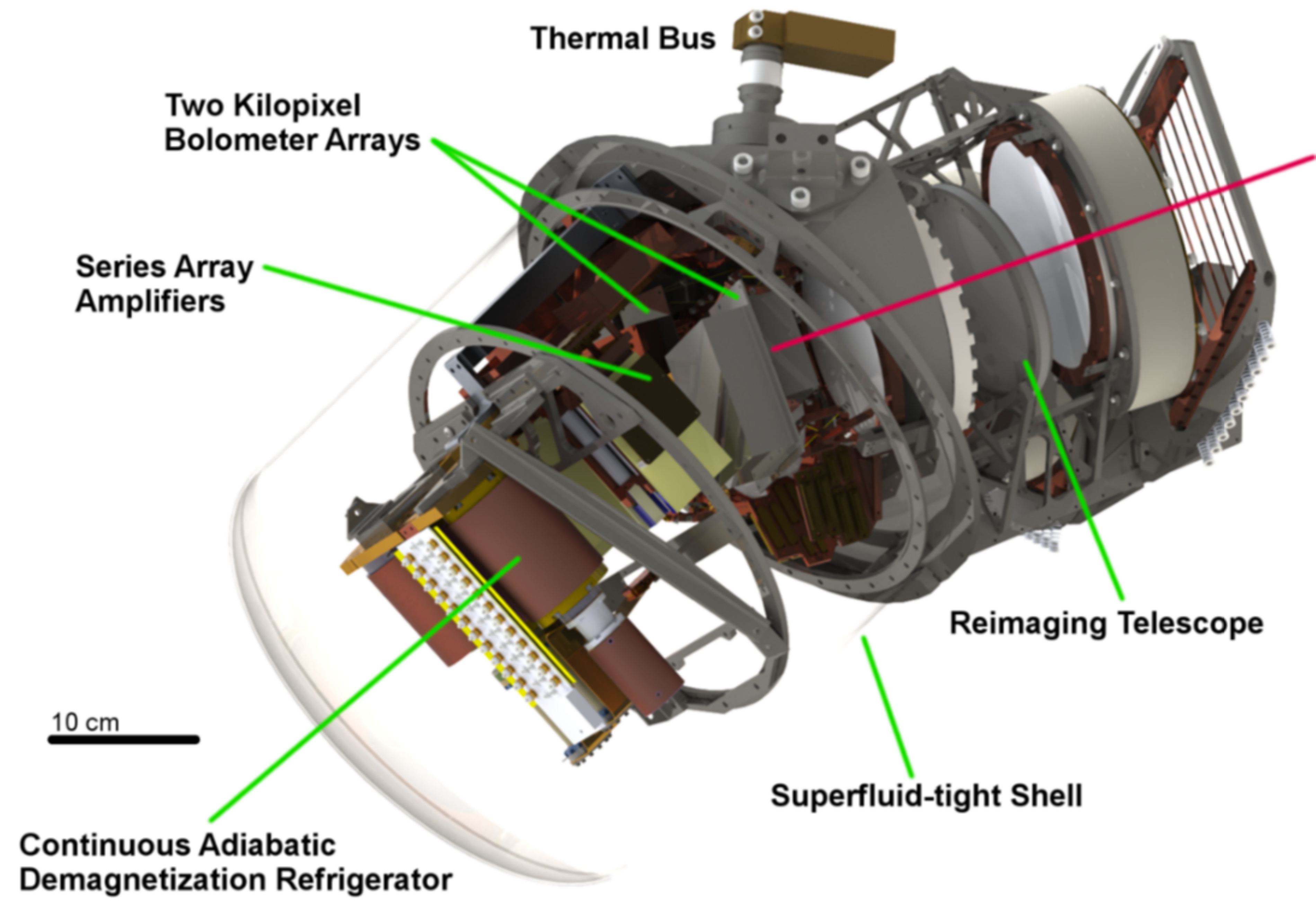}
\caption{\label{fig:piper_receiver} A section of the PIPER receiver. The red arrow indicates the path of light through the analyzer grid, cold stop, first silicon lens, filter stack, window, second silicon lens, band-defining filter, and to the $32\times 40$ TES array. The receiver serves the twin telescopes, so contains two sets of reimaging optics and both detector arrays.}
\end{figure}

% The CADR imposes stringent requirements on the superfluid tightness of the receiver. A superfluid leak can contaminate the receiver volume, and helium films can effectively short out the cold stages of the CADR. 
The receiver (Fig.\,\ref{fig:piper_receiver}) acts as a superfluid-tight enclosure within the parent dewar. Superfluid films in the receiver can spoil both the TES detector and CADR performance by shorting out thermal suspensions. Considerable effort has gone into verifying the superfluid tightness of all interfaces.

Except for a final lens, the reimaging optics reside immediately outside the enclosure's window and comprise (from the sky to detectors) a polarization analyzer grid, the cold stop, the first silicon re-imaging lens, and a radiation filter stack. The windows are fused quartz, AR-coated with Teflon\cite{2001stt..conf..410K}, and sealed against the receiver using indium. These have ${\approx}15\%$ measured loss in the $200$\,GHz band but are $1.7$\,K so do not contribute to detector loading. 
%A forthcoming publication will describe the window design, optical properties, and seal design. 

Directly inside the superfluid enclosure, following the window, a silicon lens couples light to the two $32\times 40$ TES focal planes. (The same receiver enclosure houses the detector arrays for both telescopes.) Gold-plated copper boxes house the arrays. A final band-defining filter closes the package, and epoxy loaded with stainless steel and silica\cite{2012SPIE.8452E..3IS, 2008IJIMW..29...51W} blackens the interior of the package. Two groups of 32 superconducting quantum interference device (SQUID) series array amplifiers mount to the main thermal bus, behind the lens and focal plane assemblies. Finally, the CADR is cantilevered from the lid by a drum frame, and connected by the main thermal bus to the helium bath. 

\begin{figure}
\includegraphics[scale=0.25]{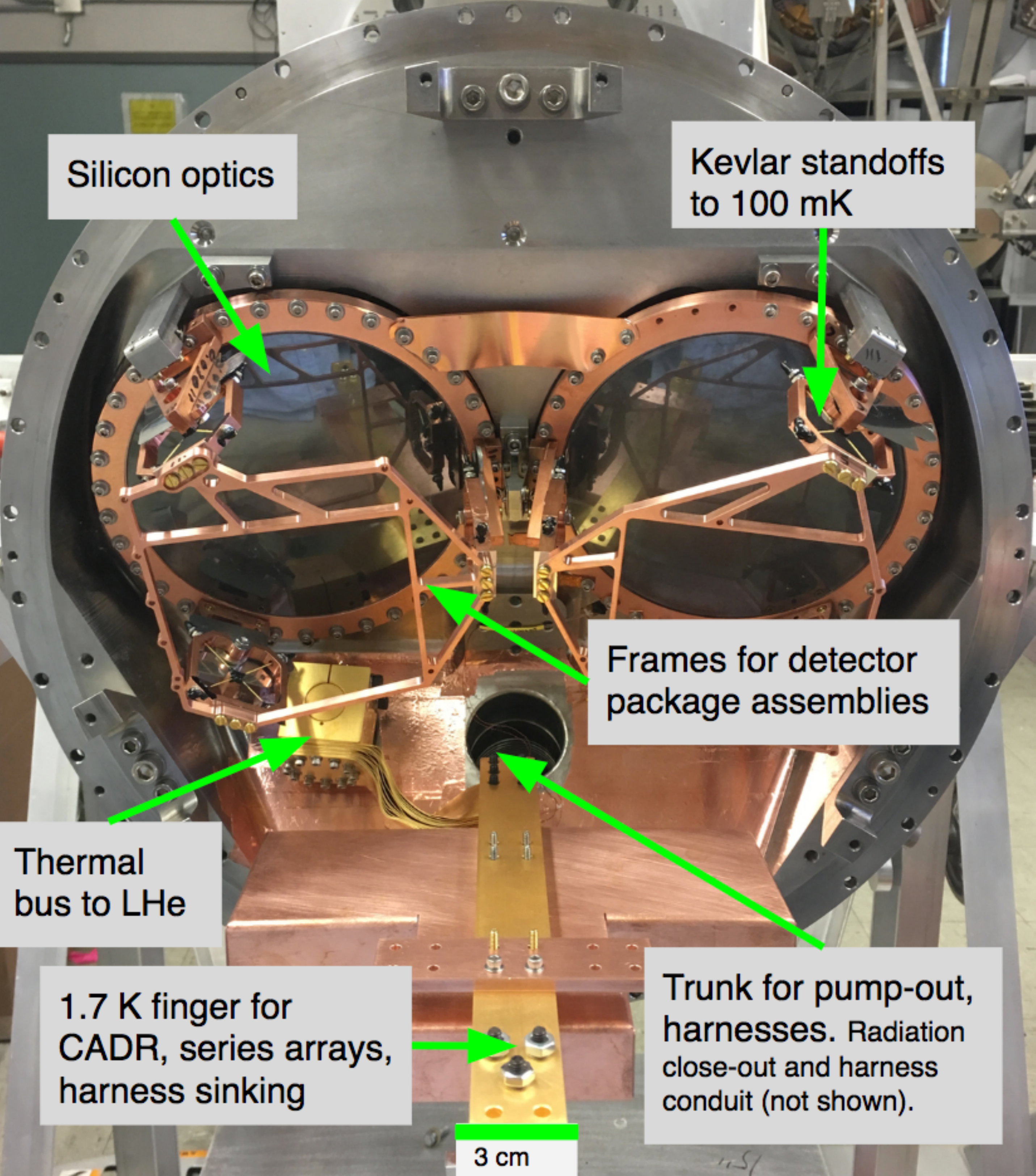}
\caption{\label{fig:receiverbus} Mechanical and thermal interfaces in the PIPER receiver.}
\end{figure}

% 19.125" is 48.58 cm
The receiver design has two practical constraints 1) the outer diameter of the receiver must be smaller than the $48.58$\,cm  inner diameter of the integration and test dewar (Sec.\,\ref{ssec:testenv}), requiring a very compact assembly, 2) the receiver lid (Fig.\ref{fig:receiverbus}) must act as the sole thermal, vacuum, optical, mechanical and electrical interface in both the test and flight cryostats. The lid must also serve as an optical alignment bench, and manage the plane change from the receiver orientation to the two optical paths. The receiver components are built outward from mechanical mount points on the lid. 

% 17.5" is 44.45 cm
% 2" is 5.1 cm
A stainless steel thin-wall bellows trunk with $5.1$\,cm ID extends from the receiver to a breakout box of feedthroughs at ambient temperature and serves as a conduit for readout and housekeeping harnesses as well as receiver pump-out. All seals must operate cryogenically and remain superfluid tight. The trunk uses $3\,\frac{3}{8}$\,in MET-SEAL flanges, and the receiver itself is $44.45$\,cm in diameter and closed out with two indium seals. 

The bellows path to the breakout box is routed to the side of the gondola to remain isolated from cryogenic gas paths on the top of the dewar. This position allows the housing to use convenient KF (Klein Flange) seals, and deploy overpressure safety, vacuum gauge instrumentation, and pump out ports that are all easily accessible on the side of the gondola and maintained at ambient temperature. High current leads for the CADR require more complex thermal accommodation (Sec.\,\ref{ssec:leads}). The vacuum manifold is well-protected and survived a landing where the gondola rolled over the top of the dewar. 

% 5 thou is 127 um
A multi-channel electronics\cite{2008JLTP..151..908B} (MCEs) module reads each of the time-domain SQUID-multiplexed TES arrays. The MCEs are mounted symmetrically about the mid-plane of the twin telescopes. Each requires 500 conductors to pass from ambient temperature to the cryogenic receiver to operate each of the two $32 \times 40$ TES arrays. The harnesses are manganin ($127\,\mu {\rm m}$ diameter) twisted pair to limit thermal conduction and are $4.5$\,m in length. The readout electronics plug directly into five MDM (Micro D-Subminiature) 100 feedthrough bulkhead plates\footnote{Vendor MPF-PI: custom part} on vacuum housings. The readout housings connect to the central break-out housing through stainless steel vacuum bellows, which also act as Faraday cages that protect the readout from radio frequency interference. 

\subsection{Heat flow in the receiver \label{ssec:heatflows}}

\begin{figure}
\includegraphics[scale=0.43]{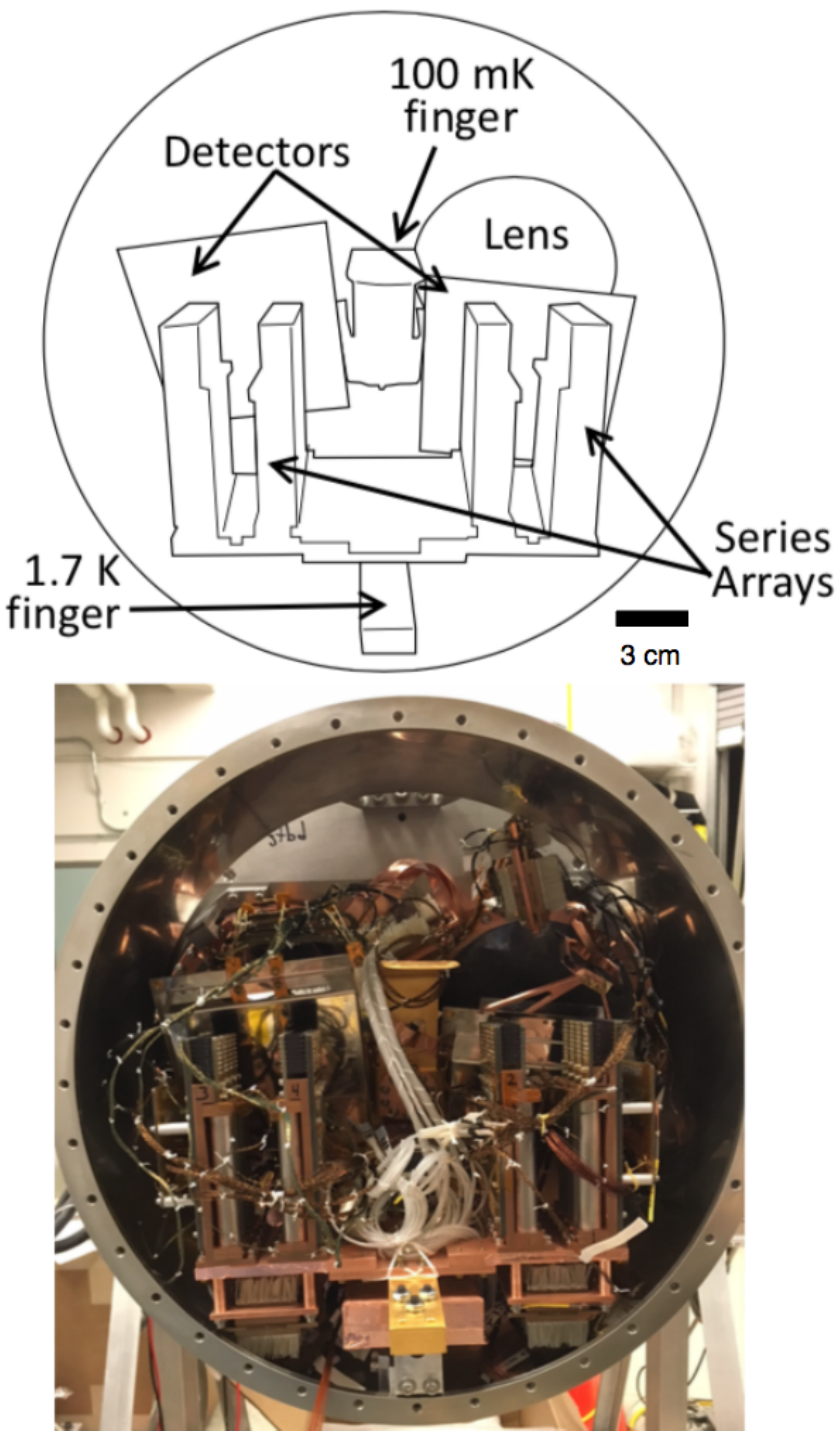}
\caption{\label{fig:piper_complete_receiver}Overview of the PIPER receiver before integration of the CADR. }
\end{figure}
% \repl{}{Fig.\,\ref{fig:receiverbus} shows the 1.7\,K finger ($2.54$\,cm wide) before installing the series arrays.}

% 7.25 is 18.4 cm
% thermal bus which is superfluid tight and $18.4$\,cm in total length. 
The receiver shell is all stainless steel to achieve CTE match with the telescope frame and dewar, but this provides poor heat conduction from the receiver interior to the large helium bath. To overcome this limited conduction, a commercial high current vacuum feedthrough\footnote{MDC 9452000, rated $3$\,kV $600$\,A} on a CF40 flange acts as a thermal bus to carry heat from the interior to the helium bath. It is $\frac{3}{4}$\,in diameter high purity copper and is superfluid tight. All thermal buses in the receiver are gold-plated\footnote{Gold plate per ASTM B488 Type III, $50-150\,{\mu}{\rm in}$, Alexandria Metal Finishers} to decrease contact resistance and maintain clean joints. Redundant super-fluid pumps fill a small reservoir clamped to the bus, fixing its temperature at $1.7$\,K. Throughout, the reference bath temperature will be $1.7$\,K, at equilibrium with the air pressure at $30$\,km minimum float altitude. The altitude is expected to remain $<37$\,km for this balloon class and mass, giving a minimum bath temperature of ${\approx}1.5$\,K.

A thermal bus of oxygen-free high conductivity (OFHC) copper connects to the feedthrough with a stack of gold-plated, diffusion-bonded and annealed OFHC copper foils. The foil bus accommodates the plane change from the feedthrough to the axis of the receiver, and also provides compliance for differential CTE between the copper bus and stainless steel frame, avoiding pressure on the ceramic seal of the feedthrough. The bus provides cooling for the lenses, series arrays (Fig.\,\ref{fig:piper_complete_receiver}), and the CADR baseplate. The readout harnesses pass from ambient temperature to 1.7\,K with no intermediate intercepts. The helium bath of the large dewar provides significant cooling power. 

\subsection{Testing and telescope integration\label{ssec:testenv}}

% 5 ft is 1.52 m
% The flight dewar has an open bore ID of $5$\,ft and requires ${\approx}5000$\,L of LHe to cool to test. Aside from the helium volume, 
It is impractical to simulate float altitude conditions in the flight dewar, as this requires both a high throughput pump and a gondola lid designed to support atmospheric pressure across the $1.5$\,m diameter of the dewar. We instead integrate the receiver and test it in a smaller system, with open bore inner diameter $48.58$\,cm and depth $1.52$\,m. The receiver is suspended by wire rope from the lid of this test dewar. The receiver windows can be blanked off, or covered with a beam-filling cold load. We use a cold load cooled with superfluid pumps, which remains isothermal to the pumped LHe bath. The test dewar uses a Rietschle-Thomas vacuum pump to simulate the conditions at float altitude by pumping on the helium bath with a high throughput of ${\approx}1000\,{\rm m}^3/{\rm h}$ at ${\sim}1\,{\rm kPa}$. A throttle valve can be used to reduce the throughput to simulate higher bath temperatures at a lower altitude.

The receiver, trunk carrying readout harnesses, and all vacuum interfaces remain {\it in situ} between testing and installation of the receiver in the telescope. Three points on the receiver lid connect to a six-point turnbuckle mount in the telescope, which allows placement relative to the optics. We use a ROMER coordinate measurement arm to align the optics and receiver form a reference axis installed in the centerline of the telescope. 

\section{$100$\,mK Cooling \label{sec:subkcooling}}

\subsection{CADR\label{ssec:cadr}}

\begin{figure}
\includegraphics[scale=0.30]{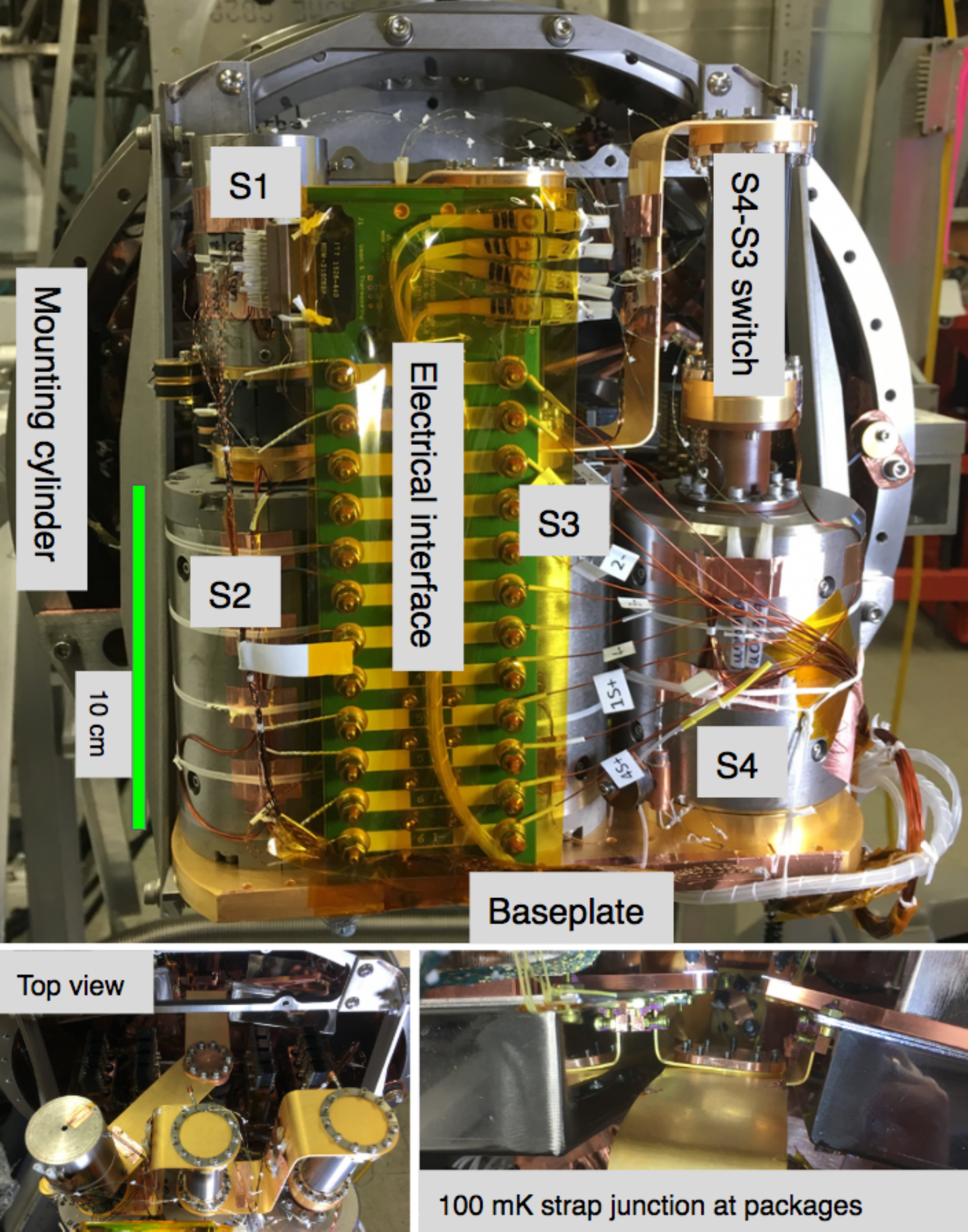}
\caption{\label{fig:CADR_overview} Overview of the PIPER CADR, which is installed as the final element of receiver integration. The lower left inset illustrates the $100$\,mK bus from the CADR to the detector packages. The lower right inset shows how this bus splits to the two detector packages.}
\end{figure}

%Name, I_max, V_max, L, B_max (vacuum), I_sat
% SCHS
\begin{table}
\caption{\label{tab:cadrpills}Properties of CADR stages 1, 2, 3, and 4 (labeled S1, S2, S3, S4). The baseplate temperature, $T_{\rm bath}$, varies from $4.6$\,K on the ground to $1.5-1.7$\,K at the float altitude. S1 provides continuous cooling at $T_{\rm S1}$, and S2 recharges S1 by providing cooling ${\approx}10$\,mK below $T_{\rm S1}$.}
\begin{ruledtabular}
\begin{tabular}{ccccc}
Stage & Mass & Inductance & $T_{\rm recycle}$ & $T_{\rm cold}$ \\
\hline
S4, GGG & 85\,g & $60.5$\,H & $T_{\rm bath}{+}0.4$\,K & $1.1$\,K \\
S3, CPA & 100\,g & $13.6$\,H & $1.35$\,K & $0.32$\,K \\
S2, CPA & 100\,g & $9.71$\,H & $0.40$\,K & $T_{\rm S1}{-}10$\,mK \\
S1, CPA & 40\,g & $3.85$\,H & - & 70-130\,mK \\
\end{tabular}
\end{ruledtabular}
\end{table}
% 3.6\,T
% $1.6$\,T
% 1\,T
% 0.16\,T

\begin{figure}
\includegraphics[scale=0.33]{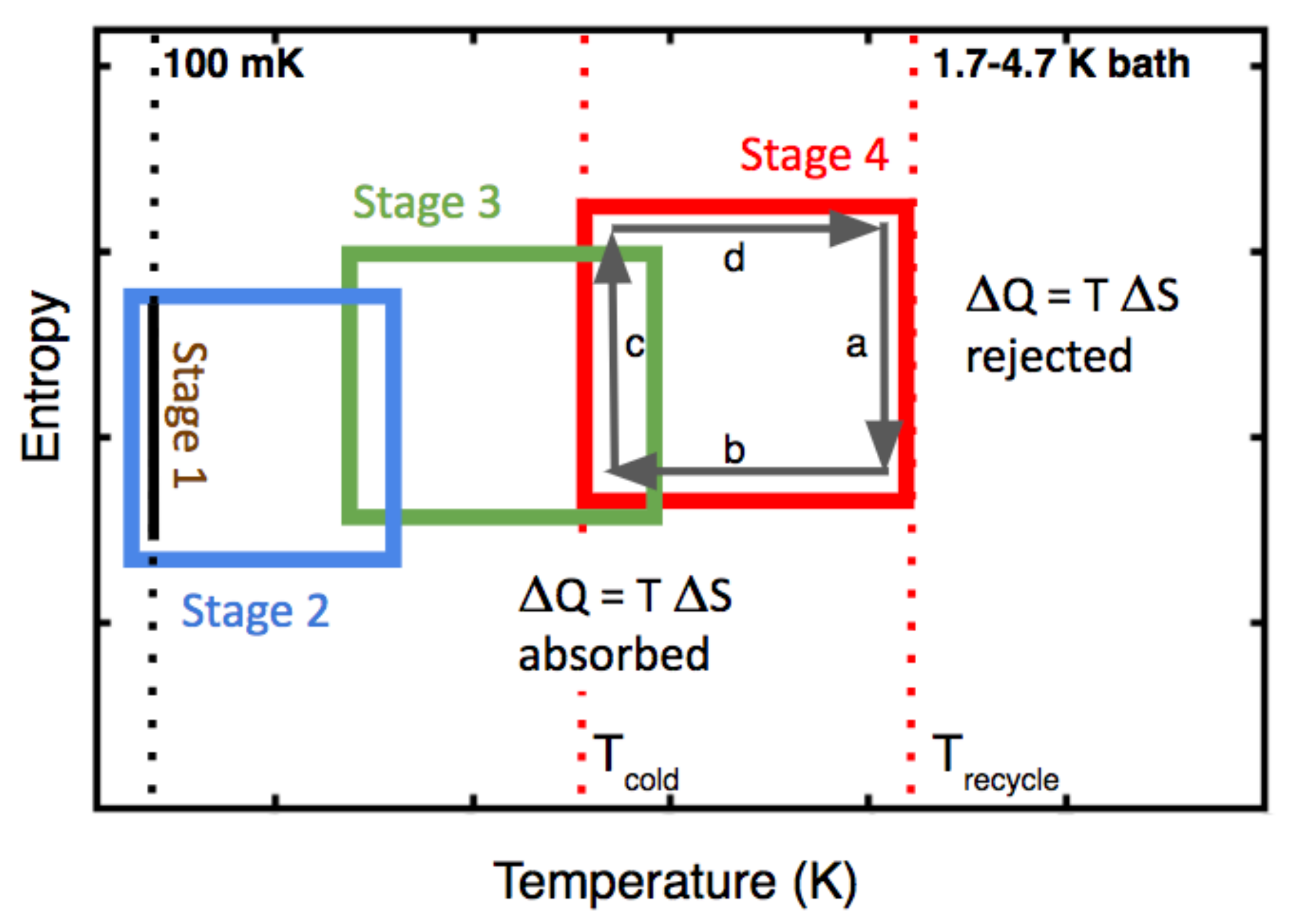}
\caption{\label{fig:stage_cycle_cartoon} Concept of multi-stage ADR operation. Each stage performs a thermodynamic cycle that rejects $T_{\rm recycle} \Delta S$ energy in isothermal magnetization against an exchange (leg {\it a}). It then decouples from the exchange and adiabatically demagnetizes to an operating point (leg {\it b}) below the exchange temperature of the next stage. In leg {\it c}, the stage absorbs $T_{\rm cold} \Delta S$. After the stage is out of cooling power (or the lower stage's recharge is complete), it decouples and adiabatically magnetizes up to $T_{\rm recycle}$ (leg d). In the continuous ADR, the coldest stage (Stage 1) simply servos at the operating point. When it is coupled to a colder Stage 2 through an active superconducting heat switch, heat flows from Stage 1 to 2. To remain isothermal, Stage 1 magnetizes, gaining cooling capacity in each cycle. Axes are not to scale. Sec.\,\ref{sec:cadroperation} and Fig.\,\ref{fig:continuous_cadr} quantitatively describe the CADR operation.}
\end{figure}

Fig.\,\ref{fig:CADR_overview} shows the CADR after installation in the receiver and Fig.\,\ref{fig:stage_cycle_cartoon} shows a cartoon of thermodynamic cycles\cite{2014Cryo...62..130S} of the four stages. Table\,\ref{tab:cadrpills} describes the physical parameters of each stage. Here GGG is gadolinium gallium garnet and CPA is chromium potassium alum\cite{2014Cryo...62..163S}. Throughout S1, S2, S3, and S4 refer to ADR stages 1, 2, 3, and 4 respectively, and BP refers to the bath baseplate of the CADR. Critical parameters in Table\,\ref{tab:cadrpills} are the temperature where the stage rejects heat to the upper stage or bath ($T_{\rm recycle}$) and the temperature from which the stage absorbs heat from the lower or detector stage ($T_{\rm cold}$). 

% 25 lb is 11.3 kg
% $9.05 \times 3.39 \times 9.3$\,in is $23 \times 8.6 \times 24$\,cm
Heritage of the system closely follows a CADR prototype developed for the Constellation-X mission concept\citep{2004Cryo...44..581S} and the architecture of the Hitomi ADR\cite{2016Cryo...74...24S, 2018JATIS...4b1403S}. The ADR crystal is suspended in the bore of a NbTi coil by Kevlar crosses at each end of the mandrel. The total envelope of the CADR is $23 \times 8.6 \times 24$\,cm, and its mass is $11.3$\,kg. 

The speed of the complete cycle that rejects heat from S1 to S2, S2 to S3, and S3 to S4 determines the CADR cooling power. The CADR design benefits from several choices that differentiate it from a single-shot ADR. First, heat switches\citep{2014Cryo...62..172D, 2017MS&E..278a2010K} must operate quickly and can tolerate higher off-state parasitic conduction to achieve higher on-state conduction for faster stage cycling. Passive switches offer not only simplicity and reliability, but they also avoid a potentially radiant getter (Sec.\,\ref{ssec:interaction}) and provide high switching speed. Second, the paramagnetic crystal must have high thermal conduction to the thermal junctions of the pill, so that the CADR may use more of the pill volume for thermal conductor interleaved into the salt\cite{2014Cryo...62..163S}. Third, faster field excursions develop higher thermal gradients between stages (reducing the cooling capacity of the pill) and operate the NbTi coils closer to regimes where they can quench. To optimize the total cooling power, each of these three design and control parameters must trade between efficiency loss per cycle and higher recycling rate. Ref.\,\onlinecite{2014Cryo...62..140S} describes additional mechanical and operational optimization.

\subsection{100 mK stage \label{ssec:100mKstage}}
\begin{figure}
\includegraphics[scale=0.27]{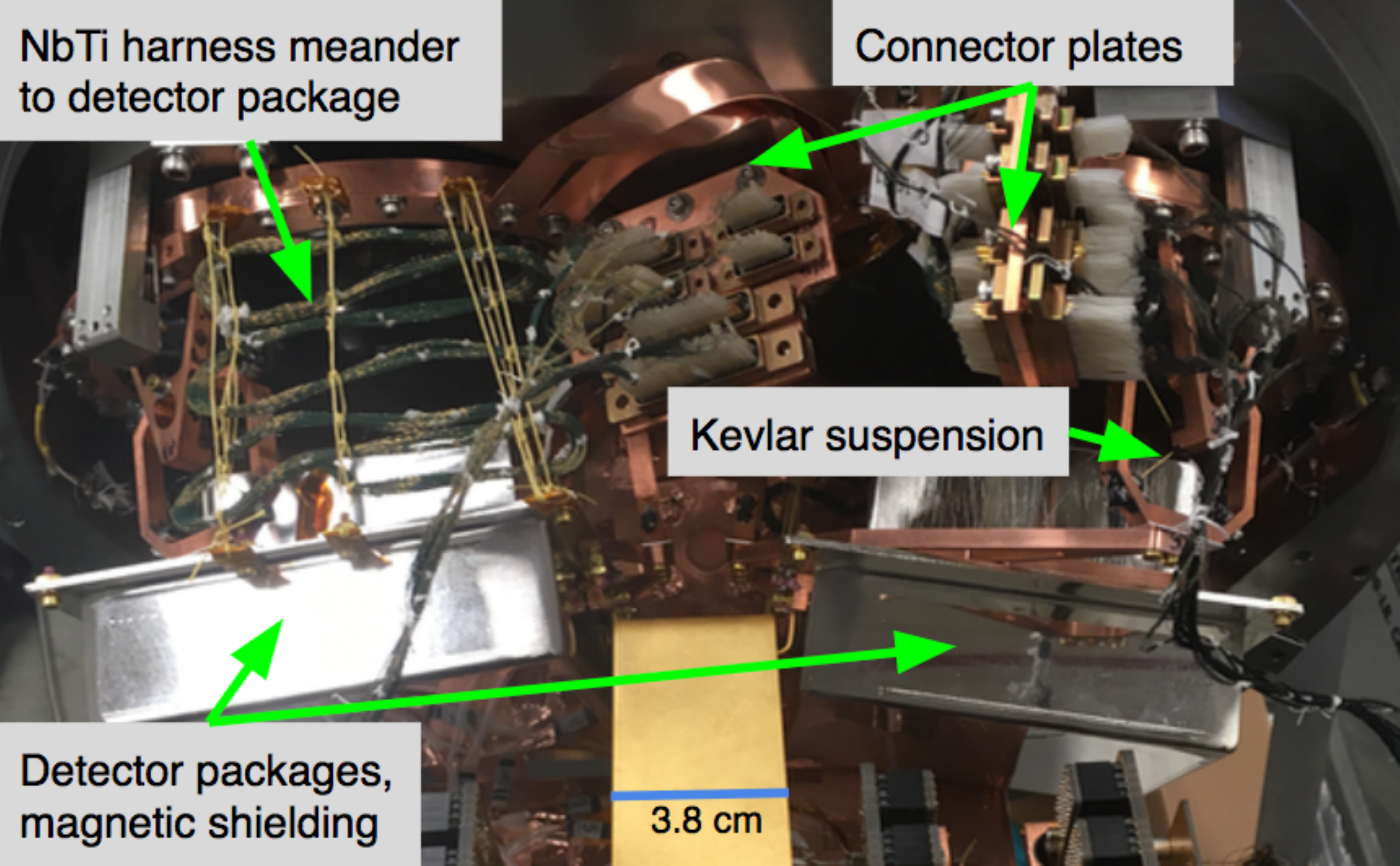}
\caption{\label{fig:100mK_suspension} Details of the $100$\,mK suspension. A meander in the cable harnesses to the packages decreases thermal conduction to the $100$\,mK detector packages from $1.7$\,K.}
\end{figure}

The detector package and lens form an integral assembly in the receiver, directly behind the window (Fig.\,\ref{fig:receiverbus}). The lens connects to the 1.7\,K bath, and the detector assembly is held up by three Kevlar suspensions (Fig.\,\ref{fig:100mK_suspension}). There is a trade-off between designing thermal stage isolation which will survive shock loading during the balloon flight, while also achieving low parasitic conduction. Following the engineering flight, the CADR had one broken Kevlar suspension and a cracked Vespel isolation shell. Both are simple to replace. A catch screw passes through the suspension to prevent the detector package from being shocked free. 

% 3 thou is 76 um
% 36 in is 91 cm
% The additional harness length in the meander reduces parasitic load on the 100\,mK stage from the harness.
The readout for each $32 \times 40$ array requires 128 conductors for the columns, 80 for the rows, and an additional 40 for 20 TES bias groups addressed in the row direction. Each conductor is $76\,\mu{\rm m}$ diameter superconducting NbTi with CuNi cladding. To reduce parasitic conduction and route the harness, the total harness length from the 100\,mK stage to the 1.7\,K lens interface is $91$\,cm and meanders across a Kevlar suspension. The harness is directly soldered to the PCB that is wirebonded to the detector array. The pigtail avoids added heat capacity and volume requirements of 16 MDM connector bodies that would be required otherwise. Sec.\,\ref{ssec:stageloading} describes loading of the $100$\,mK stage through the suspension and harness.

A silicon carrier wafer holds the detector array. The carrier is mounted in the detector package by a three-point titanium flexure that maintains the array position and takes up CTE relative to the copper package. The arrays are heat-sunk to the package through gold bonds to a diamond-turned, gold-plated boss in the package. This design\cite{2010SPIE.7741E..1QB} follows heritage from HAWC+\cite{2018JAI.....740008H}. 

% 5N 
The detector package is connected to the CADR through a copper bus\footnote{$99.999\%$ purity, ESPI metals} with four mechanical junctions. The rectangular strap sections are shaped on 3D printed forms that correspond to the interfaces in the receiver. The straps are then annealed and gold-plated. Each package has a short strap section that joins the main $100$\,mK bus. 

\subsection{Magnetic shielding and materials \label{ssec:magshield}}

Each ADR stage assembly is enclosed in a $4\%$ Fe-Si magnetic shield. When unsaturated, these maintain stray magnetic fields for each ADR stage to a level less than the Earth's field measured at the location of the detector packages. 

Due to the proximity of the detector package to the receiver window, it is not feasible to place the focal planes away from the opening of a high-aspect-ratio tube. We pursue an intermediate design where the detector package is capped from behind with an Nb box and enclosed in an Amuneal A4K shroud (Fig.\,\ref{fig:100mK_suspension}). The high purity bus feeds through a slot in the shielding.

The shields add approximately $300$\,g of cryoperm thermal mass to the $100$\,mK stage. The heat capacity of the $100$\,mK stage is at the limit of what single-shot demagnetization can cool from $2.2$\,K (worst case bath temperature) to $100$\,mK to initialize the continuous ADR operation (Sec.\,\ref{ssec:initializing}). Magnetic materials near the hybridized 2D SQUID MUX\cite{2014SPIE.9153E..3CJ} can interfere with their operation. We use brass and titanium fasteners in the copper package (avoiding stainless steel), and do not use nickel flash for gold plating. A high aspect-ratio Nb tube encloses the SQUID series array amplifiers.

A capped cylinder of $508\,\mu{\rm m}$ shielding (Amuneal amumetal) encloses the large flight dewar at ambient temperature. This shield is simulated\footnote{ANSYS Electronics Desktop, 1 Hz AC field (a conservative modulation to simulate rotation of the gondola), bounding $3\times$ larger than the shield and achieving convergence to $1\%$. The expected accuracy is $\sim 10\%$.} to suppress the Earth's field by $40\times$ and $400\times$ in the axial and transverse directions of the dewar, respectively. The large flight dewar and smaller detector package shields were not installed in the 2017 engineering flight and will be assessed in the 2019 science flight. Repeated cooldowns, shipping, and flight shocks have the potential to work-harden the shields, requiring either replacement or re-annealing depending on integrity. A Hall probe monitors shield performance. Sec.\,\ref{ssec:interaction} describes the impact of saturation of the S3 shield on the detector timestream in integrated receiver tests, which included the Nb and cryoperm detector package shielding.
%The main $100$\,mK strap connects to the CADR S1. 

\section{Continuous ADR electronics and harness\label{sec:cadrelect}}

\subsection{Control and drive electronics \label{ssec:cadrelectronics}}

Two 30U 19-inch\footnote{The unit (U) is $1.75$\,in and is a standard height in $19$\,in-wide rack frames.} rack cases hold the flight electronics and batteries. The cases protect the electronics on landing, and the front panels act as side-facing radiators. All electronics that require high power dissipation are bolted to the front panel radiators and use thermal joint compound. An Earth-IR shield reflects the radiator's view to the ground to space. 

%\begin{figure}
%\includegraphics[scale=0.4]{quiet_rack.png}
%\caption{\label{fig:quiet_rack} Overview of the housekeeping electronics. A similar 30\,U rack houses the ``noisy" electronics, such as the flight computer and one computer for the readout of each TES array.}
%\end{figure}

\begin{figure}
\includegraphics[scale=0.6]{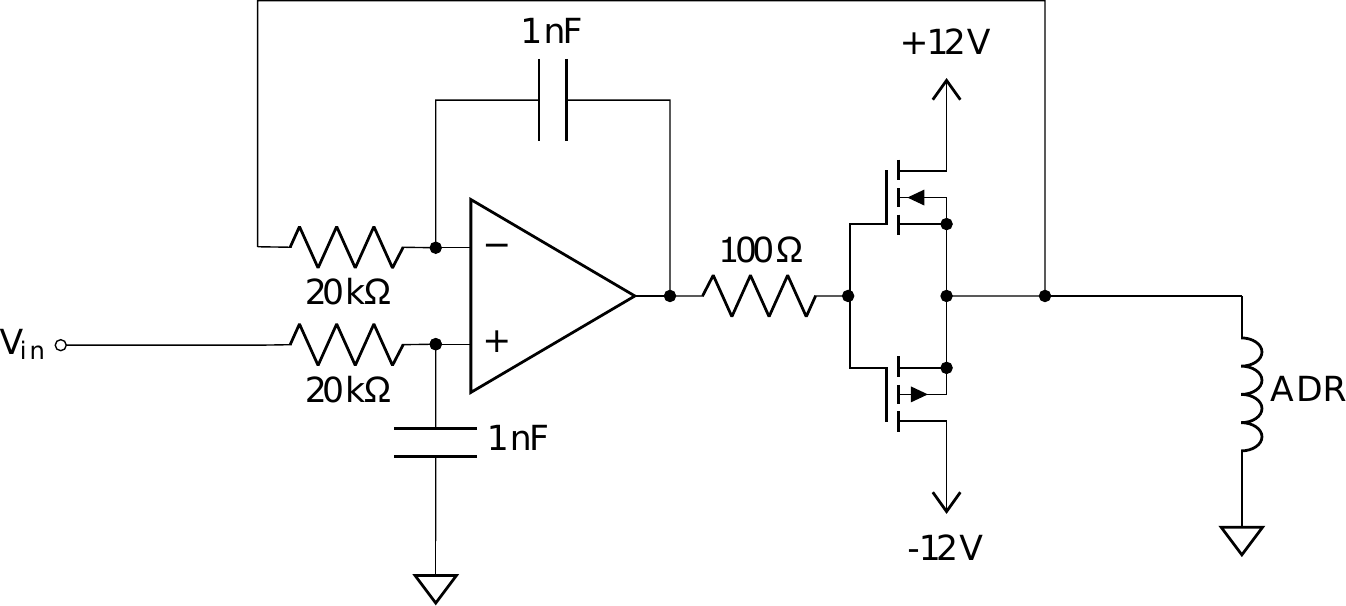}
\caption{\label{fig:boost_circuit} The high current drive circuit for the CADR coils. The current drivers use ON Semiconductor FDP047N10 N-channel power MOSFETs for the positive swing and Vishay IRF9640 P-channel power MOSFETs for the negative swing. An op-amp configured as a differential voltage integrator drives the MOSFET network's output voltage to the value set by the controller. Several supporting elements are not shown for simplicity. An input amplifier controls the bandwidth and can scale the commanded voltages from the controller if the range is insufficient. The high-current drive circuit also provides coil current monitoring using a $0.01\,\Omega$ series resistor, and coil voltage sense using a differential amplifier.}
\end{figure}

% 5/16 is 0.8 cm (Fig.\,\ref{fig:quiet_rack})
The CADR coil control electronics are split functionally into a controller and high-current drive circuit. The controller resides in a 3U subrack of backplane housekeeping cards, which communicate over fiber-optic lines to the flight computer. A dedicated 3U subrack houses the high-current drive circuits (Fig.\,\ref{fig:boost_circuit}) and has a $0.8$\,cm thick aluminum baseplate for heat sinking. 

The controller output actuates the voltage at the output of the high-current source and accepts commands to maintain several possible constants: 1) voltage on the coil's voltage taps, yielding a constant current rate ($dI/dt$), 2) resistance of the ruthenium oxide thermometry on a stage for isothermal control and 3) a constant coil current. Note that a constant voltage from the drive circuit does not correspond to a constant voltage across the coil due to line resistance as an LR circuit, so some active control is required. A microcontroller runs the proportional integral derivative (PID) control loop at ${\approx}8$\,Hz and 16-bit output resolution. Software services on the flight computer manage the high-level operation of the cycle (Sec.\,\ref{ssec:cadrcontrol}) and the firmware on the PID controller cards servos in response to flight computer commands. The PIPER electronics have heritage from several previous missions in the NASA Goddard balloon group\citep{2008SPIE.7020E..2JH}.

% with $54$\,nA typical square wave excitation current
All CADR thermometers are the Lakeshore RX-102A-CD, which we purchase uncalibrated, and calibrate in a dilution refrigerator against a Lakeshore reference. A 16-bit analog--to--digital converter\footnote{Vendor Analog Devices: AD7685} reads the ruthenium oxide thermometry in a bridge configuration with heritage from ARCADE \citep{2002RScI...73.3659F}. For simplicity and interchangeability, the controller retains no knowledge of the calibration from resistance to temperature. The flight software (Sec.\,\ref{ssec:cadrcontrol}) converts temperature setpoints to resistance setpoint commands using calibration curves particular to each thermometer. The controller is effectively linear in temperature for small fluctuations around the setpoint, but the PID parameters must be chosen to give stable behavior over the operational resistance range.

Batteries must power the CADR with sufficient capacity for operation throughout the mission, estimated to be at most $24$\,hours. While lithium sulfur dioxide (SAFT G62 series) batteries power the rest of the flight electronics, we use rechargeable, sealed lead-acid batteries for the CADR due to concern for catastrophic failure of the lithium batteries in high-current quench events. Two cells\footnote{Vendor Power-Sonic: PS-121000B} in parallel ($200$\,Ah total) supply $+12$\,V. Current from the $-12$\,V supply is infrequently needed, and a smaller\footnote{Vendor Power-Sonic: PS-12350B} battery ($35$\,Ah) is sufficient. For initial cooldown, the drivers must magnetize all stages and provide $8.5$\,A. In continuous operation, current requirements are typically $4.5$\,A, and peak at $6.5$\,A.

% 15 feet ambient and 10 feet cryogenic
%Several areas required significant development work. An early design of the high-current driver used a Texas Instruments OPA549T power op-amp with a $\pm6$\,V supply. We found that these drivers had limited lifetime, likely due to coil quench events damaging the power op-amps and despite having clamp diodes from the output terminals to the power supplies. 
To be able to supply the maximum of $4$\,A per coil with a FET driver range of $\pm 8$\,V, the resistance of the harness to the coils must be $<2\,\Omega$. In PIPER, the harness is $4.5$\,m at ambient temperature, and $3$\,m cryogenic, necessitating the use of $18$\,AWG copper wire for the harness to the CADR. Finally, the grounding design must avoid returning large coil currents through the common ground reference point for the sensitive PID control electronics. 

\subsection{Cooling the high-current leads\label{ssec:leads}}

\begin{figure}
\includegraphics[scale=0.35]{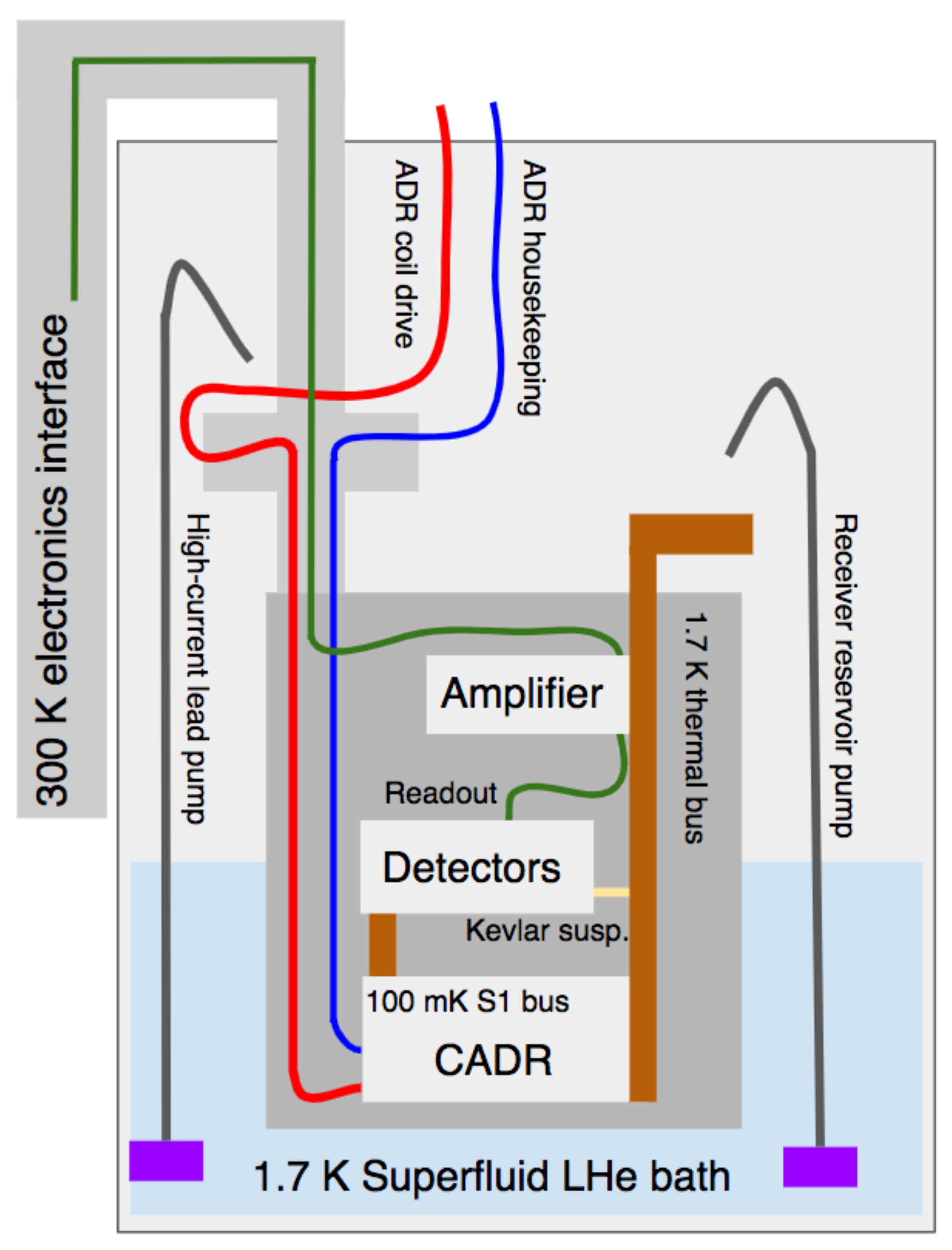}
\caption{\label{fig:thermal_diagram} Cartoon of the PIPER receiver emphasizing connections of the harnesses and thermal paths.}
\end{figure}

High current copper leads run from ambient temperature to the 1.7\,K bath of the receiver with no other available temperature intercepts, so must be normal metal rather than employ, for example, high-temperature superconducting breaks. Copper with excellent electrical conductivity necessarily has excellent thermal conductivity from room temperature. A run of the high-current leads through the vacuum trunk to the receiver produces thermal loading in the receiver that overwhelms the thermal bus (Sec.\,\ref{ssec:heatflows}) to the helium bath. 

Helium vapor and liquid in the main dewar space pre-cool the high current CADR leads, which then enter the receiver through an 8-pin \footnote{Vendor MPF-PI: A0757-1-CF} superfluid-tight feedthrough (Fig.\,\ref{fig:thermal_diagram}). We solder the high current leads directly onto both the inside and outside pins of the feedthrough, effectively eliminating contact resistance and producing a mechanically robust connection. The high current lines connect to the feedthrough assembly in a cryogenically-compatible circular connector bulkhead to avoid a captive cable. While the feedthrough and receiver are submerged in LHe before the flight, at float altitude, the feedthrough is approximately ${>}60$\,cm above the helium bath. The connector and feedthrough itself are cooled with redundant superfluid LHe pumps, mitigating any Joule power. Inside the receiver, $0.31$\,mm diameter NbTi wire\footnote{SUPERCON SC-T48B-M-0.5mm} with $0.5$\,mm copper cladding transmits the high current to the CADR. The copper cladding provides high thermal and electrical conduction in case of a quench or normal metal conduction, making it more robust than the $76\,\mu{\rm m}$ diameter NbTi (CuNi cladding) used elsewhere. Joule power in the high-current harness produces a $30$\,mK increase in the baseplate temperature during CADR operation. In comparison, heat rejection during the S4 cycle increases the baseplate temperature by ${\approx}45$\,mK.
%Fig.\,\ref{fig:high_current_lead_splice} shows the high current feedthrough spliced onto the CADR high-current coil leads in a cross junction in the trunk bellows. 

%\begin{figure}
%\includegraphics[scale=0.35]{high_current_lead_splice.png}
%\caption{\label{fig:high_current_lead_splice} Splicing the high-current superfluid-tight feedthrough onto the CADR leads inside the trunk to the receiver submarine.}
%\end{figure}

%To maintain a unified CADR control harness, 
The thermometry and voltage tap channels also follow the high current lines through the helium dewar and enter the receiver through a superfluid-tight 41-pin\footnote{Vendor CeramTec: 24016-01-KF} circular connector feedthrough. Both feedthroughs are mounted in a cross in the trunk, directly above the receiver (Fig.\,\ref{fig:thermal_diagram}).

\subsection{Electrical Interfaces \label{ssec:cadrelecinter}}

A custom printed circuit board (PCB) provides a robust and repeatable electrical interface to the CADR (Fig.\,\ref{fig:CADR_overview}). An MDM31 connector carries the voltage taps and stage thermometry. Thermometry instrumented on the CADR stages is read over $76\,\mu{\rm m}$ CuNi-clad NbTi to limit parasitic conduction. All junctions of the high current leads are potential sites of single-point failure. The system must be robust to multiple integration cycles, vibration during shipment, and (as much as possible) to parachute and landing shock. Each ADR stage has a primary and redundant thermometer.

High current junctions use gold-plated copper bobbins\footnote{Vendor Lakeshore: HSB-8} to which copper-clad NbTi magnet wires are wrapped and soldered. The trace size on the interface board is maximized to sink heat from the high current channels to the board groundplane and to minimize Joule heating. The bobbins are attached to the PCB with $\#4$ fasteners and three Belleville washers. To prepare the flight instrument, we additionally stake the fasteners to maintain the junction under vibration in shipping. 

\section{CADR operation\label{sec:cadroperation}}

During the 2017 engineering flight, the S4 coil current harness opened intermittently, consistent with an air-side connector of the high-current CADR feedthrough that was not cryogenically compatible (Sec.\,\ref{ssec:leads}). This open-circuit appeared after shipping, and was not serviceable before the flight but was subsequently remedied by soldering directly onto the feedthrough (Sec.\,\ref{ssec:leads}). During the flight, S3, S2, S1, and the SCHS performed nominally in single-shot operation. CADR data used throughout are from integrated receiver testing at NASA Goddard. These tests use the flight electronics, software, and commanding identical to that available in the ground station. We throttle the pumped bath (Sec.\,\ref{ssec:testenv}) to simulate a range of CADR base temperatures equivalent to those expected in the balloon flight.

\subsection{Stage controller logic and architecture \label{ssec:cadrcontrol}}

\begin{figure}
\includegraphics[scale=0.55]{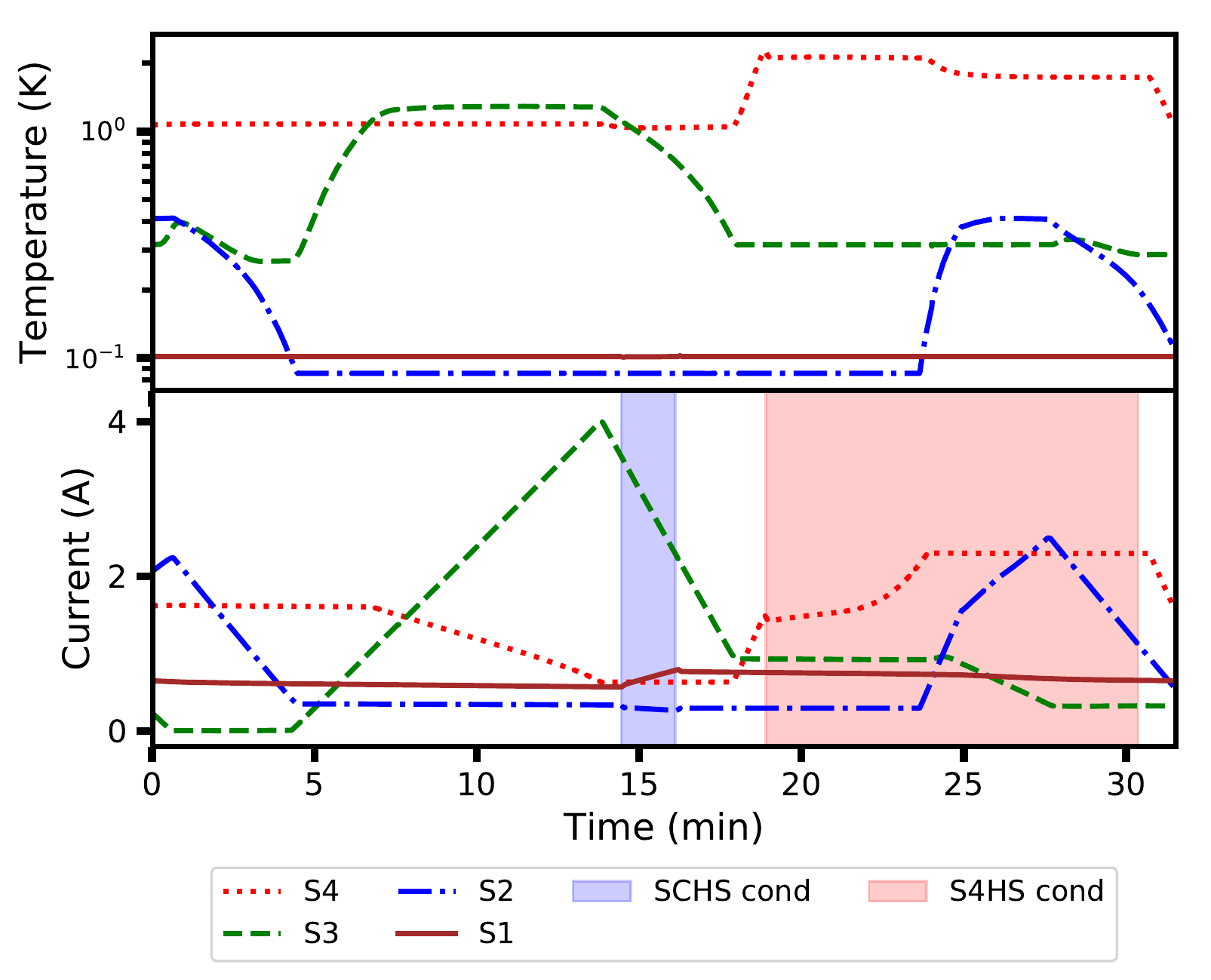}
\caption{\label{fig:continuous_cadr} One typical cycle of continuous operation of the ADR, illustrated with data from integrated receiver testing at NASA Goddard. S1 remains at $100$\,mK, with heat cycled out through stages 2, 3, and 4, with S4 ultimately rejecting heat to the pumped helium bath.}
\end{figure}

Fig.\,\ref{fig:continuous_cadr} shows the ADR coil currents and temperatures for continuous operation.  Each stage controller has the same underlying logic but has different parameters for the operating points (Table\,\ref{tab:cadrpills}). Separate, asynchronous software processes on the flight computer control the ADR stages and communicate through a \verb+redis+ database\footnote{Redis database: redis.io}. 

\begin{table}
\caption{\label{tab:cadrswitches}Properties of the heat switches between each of the CADR stages. The S2-S1 connection is a superconducting heat-switch (SCHS) that employs a lead conductor.}
\begin{ruledtabular}
\begin{tabular}{ccc}
Connection & Condition & Type \\
\hline
BP-S4 & $T_{\rm getter} < 5$\,K & Active gas-gap \\
S4-S3 & $T_{\rm S3} < 1000$\,mK, off by $T_{\rm S3} < 750$\,mK & Passive gas-gap \\
S3-S2 & $T_{\rm S2} < 300$\,mK, off by $T_{\rm S2} < 130$\,mK & Passive gas-gap \\
S2-S1 & $I_{\rm SCHS} < 200$\,mA, $T_{\rm S1} < 250$\,mK & Active SCHS \\
\end{tabular}
\end{ruledtabular}
\end{table}

Heat switches\citep{2014Cryo...62..172D} control the heat flow between ADR stages throughout the cycle. Table\,\ref{tab:cadrswitches} shows the properties of the heat switches connecting the four stages and baseplate. Between S4 and the baseplate, an active gas-gap switch\citep{2015MS&E..101a2157K} allows for flexibility in the temperature at which S4 couples to the bath. This flexibility is essential to manage variations in the bath temperature with altitude. In contrast, the S2-S3 and S3-S4 heat switches are passive gas gap switches\citep{2017MS&E..278a2010K} that make a transition from conducting to non-conducting in a narrow band of temperatures. Below their critical temperature, gas adsorbs on the cold side of the switch, effectively shutting down gas conduction. At higher temperatures, the gas desorbs and can effectively conduct. This yields rapid switching with no electrical input. Faster switching improves the CADR cooling power. Passive switches require a heat exchange at a fixed temperature, but this is well-matched to thermodynamics of the S2 and S3 cycles. From test data, we find off-state parasitic conduction in the S3-S2 and S4-S3 passive switches to be $3\,\mu{\rm W/K}$ and $7\,\mu{\rm W/K}$, respectively. All gas-gap switched employed here use $^3{\rm He}$.

An active superconducting heat switch (SCHS) connects S2 and S1. Here a section of superconducting lead is a poor thermal conductor in the off state. An applied field with small Helmholtz coil drives the lead normal and into thermal conduction. This allows for rapid control of the exchange between S2 and S1 across a range of temperatures (up to ${\sim}250$\,mK), providing flexibility in the continuous stage temperature and timing of the exchange. From test data, we find off-state parasitic conduction of $36\,\mu{\rm W/K}$.

Starting with the recycling operation, the controller first checks to see that the intercept stage is ready, providing a temperature slightly below $T_{\rm recycle}$. Here ``intercept" refers to either the CADR baseplate (for S4) or the ADR stage that provides cooling. For example, S4 is the intercept for S3, and referring to Table\,\ref{tab:cadrpills}, S3 will wait until S4 is at $1.1$\,K before starting to recycle. At this point, the S3 controller seeks to maintain the S3 temperature at $T_{\rm recycle}(S3) = 1.35$\,K, $\Delta T = 0.25$\,K above the S4 intercept. Heat flow to the intercept stage across $\Delta T$ requires magnetization of S3 and demagnetization of S4. The magnetization will continue until either the maximum current (hence magnetization) is reached, or the intercept is out of cooling capacity.

%The next step of the state controller is responsible for adiabatic demagnetization. 
At this point, the stage has reached its target magnetization, and the controller can start adiabatic demagnetization to the operating temperature, $T_{\rm cold}$. If the stage conducts to its intercept through an active switch, the controller will command the switch to become nonconducting. In the case of a passive gas gap switch, the stage will initially conduct to its intercept until dropping low enough that the gas gap becomes non-conducting. Once cold, the controller sets a flag that allows the next stage controller to see that its intercept is ready.

While the cyclic operation of S4, S3, and S2 is all autonomously controlled, we leave the S2-S1 interchange under manual control. Manual control comprises 1) setting the S1 PID controller at the desired detector operating temperature and 2) commanding the SCHS. SCHS control allows the operator to recharge S1 as needed and avoid interruptions during critical operations, such as detector biasing.

The stages must operate robustly through their thermodynamic cycles with limited input from the ground. In conventional balloon flights, the uplink is limited to commands with one-byte address and one-byte value, and the downlink telemetry is 57.6\,kbaud. Commands from the ground perform the initial demagnetization (Sec.\,\ref{ssec:initializing}) to the continuous state and then initiate the S4, S3, and S2 controller state machines, which operate autonomously. 

%In addition to the stage logic described in Sec.\,\ref{ssec:cadrcontrol}, practical CADR operation must account for several other considerations, described here, in turn.

\subsection{Initializing the continuous state \label{ssec:initializing}}

\begin{figure}
\includegraphics[scale=0.55]{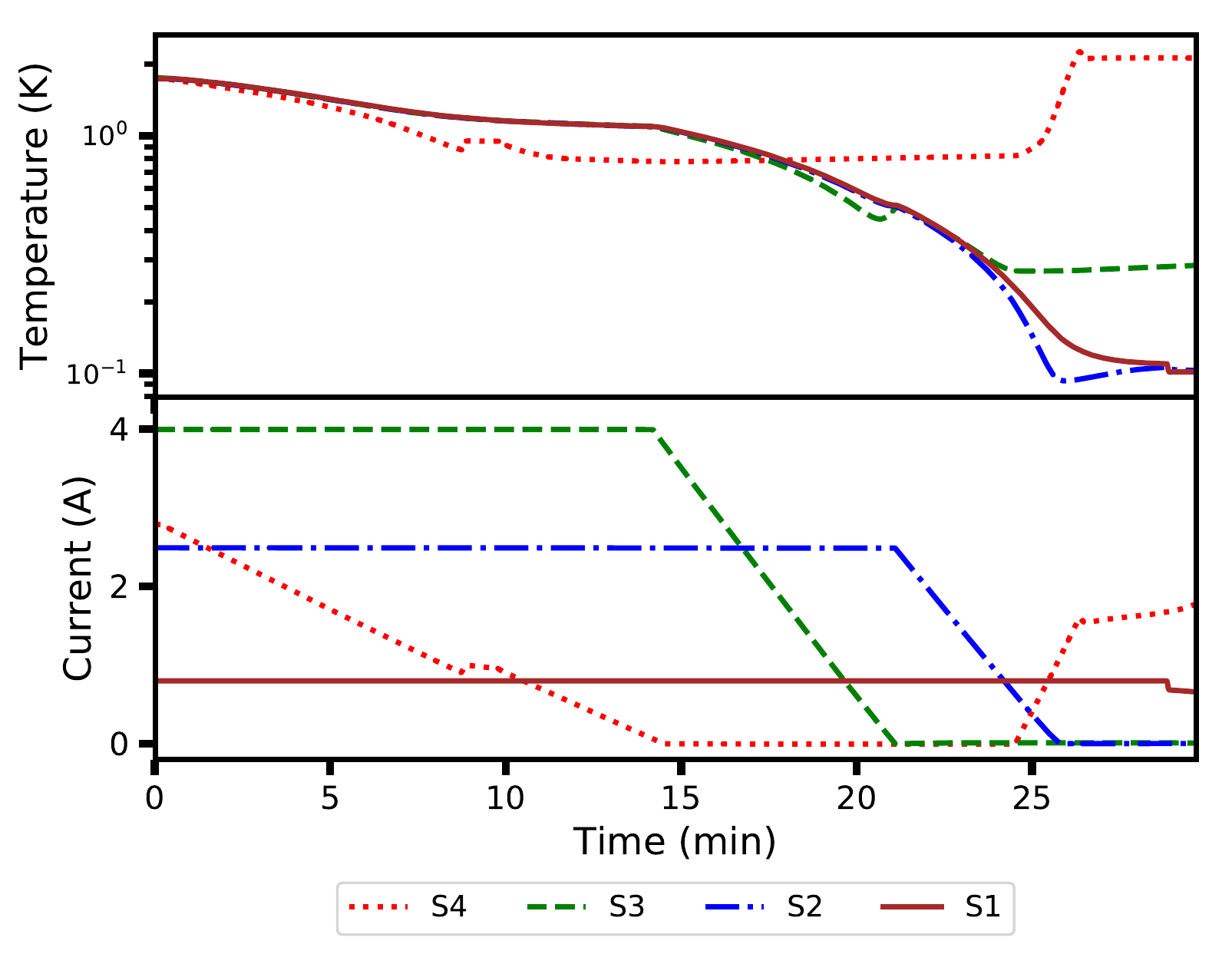}
\caption{\label{fig:initializing_cadr} Preparing continuous operation by sequentially demagnetizing the stages. Once S1 reaches $100$\,mK, S4 re-magnetizes. It proceeds to recharge S3, then S3 recharges S2, and S2 recharges S1. Data are from integrated receiver testing at NASA Goddard.}
\end{figure}

The CADR stages and detector packages start at the CADR baseplate temperature ($1.7-4.7$\,K), so are far from their desired operational temperature ranges and have high heat capacity. Fig.\,\ref{fig:initializing_cadr} shows the progress of the initial demagnetization. To begin the cycle, we magnetize S4 to $2.8$\,A, S3 to $4$\,A, S2 to $2.5$\,A, and S1 to $800$\,mA with the 4-BP switch coupled to the bath. Once the S4 is nearly in equilibrium with the bath temperature, we decouple the 4-BP switch and demagnetize S4 to a temperature of $1.1$\,K. This cooling pulls S3, S2, S1, and the detectors to $1.1$\,K. S4 cannot pull the upper stages much below this temperature because the passive 3-4 switch conduction drops rapidly (Table\,\ref{tab:cadrswitches}). We then demagnetize S3 to 0 A, reaching $550$\,mK from a $2$\,K base plate. From here, we demagnetize S2 and reach 100 mK. After S2 has pulled S3 below $300$\,mK, we start the S4 cycle state machine. Once S2 is demagnetized, S1 can start continuous operation. 

\subsection{Ramp rates \label{ssec:optimalrates}}

The field ramp rate must be optimized to account for a range of conditions, principally to minimize thermal gradients while also achieving a high cycle rate. Each stage in the CADR has three typical cooling profiles that each have optimal ramp rates:
1) cooling downstream stages during initialization (Sec.\,\ref{ssec:initializing}). Here the stage is cooling the high heat capacity of subsequent stages and must ramp relatively slowly at to remain within ${\approx}80\%$ of the lower stage temperature  (Fig.\,\ref{fig:initializing_cadr}), requiring $15$\,min, $7$\,min and $4$\,min for demagnetization of S4, S3, and S2, respectively. The heat capacity drops as the $100$\,mK stage cools from $1.7$\,K, allowing progressively faster operation, 2) ramping between isothermal exchanges. Here a stage decouples from other stages and can magnetize or demagnetize quickly to change temperatures, requiring approximately $5$\,min for each stage to demagnetize, and 3) isothermal exchanges between ADR stages during recycling operations of S4, S3 and S2 are $5-10$\,min. Fig.\,\ref{fig:continuous_cadr} shows the latter two ramp periods. The system has excess cooling power (Sec.\,\ref{ssec:stageloading}), so we have chosen conservative rates which are far from a regime where the magnets can quench.
%In practice, these ramp rates are chosen to be the fastest possible which will not quench the coil, \red{XYZ}\,A/m. 

\subsection{Variations in operational envelopes \label{ssec:openvelope}}

The changes in pressure during the flight produce a range of LHe bath temperatures. During the ascent, the falling pressure pumps the LHe bath and the temperature drops from $4.3$\,K on the ground to as low as $1.5$\,K at balloon float. The superfluid pump system cannot operate until the LHe bath is below 2.17\,K, achieved at ${\approx}24$\,km altitude. After reaching float altitude, a zero-pressure balloon can sag in altitude overnight, causing the bath temperature to increase. The engineering flight had more sag than expected and resulted in a $0.1$\,K bath increase. At nominal float altitudes $30-37$\,km, the CADR baseplate runs between $1.5-1.7$\,K. Tests of the integrated receiver at NASA Goddard verified operation of the CADR up to the worst-case flight conditions of a $2.4$\,K baseplate. 

The CADR must have robust, primarily autonomous operation across the baseplate temperature range with minimal cycle parameter changes. We use the significant spare capacity of the S4 to take up these differences in CADR baseplate temperature. S3, S2, and S1 all operate at their nominal parameters regardless of the baseplate temperature, as they must to continue to work with passive switches at fixed transition temperatures. S4 uses a GGG rather than CPA (as is used in S3, S2, and S1), and can operate from ${\sim}700$\,mK to $5.8$\,K. S4 provides significantly more cooling capacity than is needed at float altitude. 

\subsection{Steady-state operation \label{ssec:steadystate}}

The CADR has approximately $5\times$ the necessary cooling power at float altitude conditions (Sec.,\ref{ssec:thermodynamics}). In the auto-cycle state machine (Sec.\,\ref{ssec:cadrcontrol}), unused capacity results in stage currents that rise with each complete cycle of the CADR. S2 and S1 reach their maximum currents on each cycle, shortening the S3 cycle. Although the cycle reaches a steady-state, we choose to park the auto-cycle with stages at their cold temperatures when there is some spare cooling capacity (e.g., the build-up of excess current on S2 and S1). This 1) minimizes parasitic conduction through the heat switches, 2) reduces variations of the magnetic fields (reducing detector interaction, Sec.\,\ref{ssec:interaction}), and 3) uses less battery power. 

\subsection{Quenches \label{ssec:quenches}}

Quenches can be caused by magnetizing or demagnetizing at high rates, over-magnetizing a coil, or by operating a coil at temperature $>5.2$\,K. Default parameters and working limits in the software controller and firmware prevent all three quench conditions. The ADR coils have protection diodes, and all quenches to-date have produced thermal transients, but no coil damage. It takes nearly two hours to recover from a quench (including stabilization, re-magnetization, and demagnetization to continuous operation at 100 mK).

The S4 coil initially quenched below its required $2.8$\,A operating current and must be trained up to its operating range\cite{2016ITAS...2614598V}. This behavior is believed to be due to a mechanical defect (running in microcracks in the coil potting) in the specific coil, which will be replaced in subsequent work. Training requires six quenches over $45$\,min in the cryogenic commissioning of the receiver. Provided the magnet remains at cryogenic temperatures, no additional quenches before launch are required. Coils for the S1, S2, S3, and SCHS do not require re-training and are only exercised up to $100$\,mA in the pre-flight check. 

\subsection{Comparison to single-shot ADR systems}

The CADR provides several advantages over single-shot operation aside from a continuous cold stage temperature. A single-shot ADR stage requires the capacity to cool for the duration of science operation, here $>12$\,hr. In contrast, a continuous stage must be sufficient to hold for $<20$\,min of a single-stage cycle (Fig.\,\ref{fig:continuous_cadr}), yielding a reduction in salt volume and mass, which propagates to the coil and shield volumes \citep{2004Cryo...44..581S}. For example, in cryogen-free mode from $4.5$\,K, the 3-stage ADR in Astro-H\citep{2016Cryo...74...24S} has a hold time of 15\,hr for $<1\,\mu{\rm W}$ load at $50$\,mK, and weighs $15$\,kg. Analogous operation of the CADR provides $>4\,\mu{\rm W}$ cooling with less mass. A multistage design also provides flexibility to accommodate multiple bath temperatures (through different salt types) and to reduce parasitic power on cold stages\citep{2007JLTP..148..915S}.

Single-shot, single-stage ADRs must limit parasitic conduction to achieve long hold times, so benefit from mechanical heat switches \cite{1995Cryo...35..303H} or backing coolers \cite{2006RScI...77g1101R}. In contrast, lower inter-stage parasitic power in the multistage CADR allows for the use of gas-gap switches, which avoid mechanisms, or any electrical input (in the case of passive\cite{2017MS&E..278a2010K} switches). Low parasitic loading also means that the coldest stage can be relatively small and low inductance (Table.\,\ref{tab:cadrpills}), making it easier to control. All stages are driven by identical PID current controllers (Sec.\,\ref{ssec:cadrelectronics}) and stage control logic (Sec.\,\ref{sec:cadroperation}), which mitigates the added complexity of multiple stages. Sec.\,\ref{ssec:cadrelectronics} describes the power requirements, which are comparable to the mean of a single-shot stage with $9$\,A peak magnetization. Spreading the cooling power over many short cycles also levels loading in cryocooler systems \cite{2014Cryo...62..140S}, making it of interest for ultra-long duration ballooning or space applications.

\section{Performance\label{sec:performance}}

\begin{table}
\caption{\label{tab:cadrparams}Measured operational parameters of the CADR. A range of parasitic loading is given for when the intercept stage (e.g., S2 for S1) is at a higher temperature recharging. The difference in power represents the parasitic loading across the heat switch. The charge ratio is the ratio of current gained in the magnetizing stage to current lost in the demagnetizing stage at each isothermal exchange (S1 charges no stage, so no ratio is given).}
\begin{ruledtabular}
\begin{tabular}{ccccc}
Stage & $\Delta T/\Delta I$ & Charge ratio & $P_{\rm parasitic}$ & $I_{\rm max}$ \\
\hline
S4 & 1\,mK/mA & 3.3 & $40\,\mu{\rm W}$ & $2.8$\,A \\
S3 & 0.31\,mK/mA & 2.1 & $(3{-}10)\,\mu{\rm W}$ & $4$\,A \\
S2 & 0.15\,mK/mA & 4.3 & $({<}1{-}4)\,\mu{\rm W}$ & $2.5$\,A \\
S1 & 0.1\,mK/mA & $-$ & $(2{-}13)\,\mu{\rm W}$ & $0.8$\,A 
\end{tabular}
\end{ruledtabular}
\end{table}
% also found 0.16\,mK/mA at a different time
% 0.17\,mK/mA appears to be true rate for S2, but there's some loss from S2/S3 coupling, so 0.15

\subsection{$100$\,mK stage loading\label{ssec:stageloading}}

% Conversion from current to magnetic field is roughly $0.2$\,T/A for S1, $0.4$\,T/A for S2 and S3, and $1.3$\,T/A.
% 14 thou is 356 um diameter
% 21" is 53 cm, 3.3 in is 8.4 cm
We infer loading of $2.1\,{\mu}W$ on the $100$\,mK stage (Table\,\ref{tab:cadrparams}), calibrated by applying a known power to a heater and measuring the difference in the S1 coil current ramp rate at a fixed temperature. We estimate the NbTi readout harness loading from both arrays to be $610$\,nW, and the Kevlar suspensions to be $1.1\,{\mu}{\rm W}$, giving $1.7\,{\mu}{\rm W}$ in total. The harness from the $100$\,mK detector package to the lens traverses a $53$\,cm meander before being anchored at the lens at $1.7$\,K. The three Kevlar suspensions for each package have four $16$\,mm links, each of $356\,\mu{\rm m}$ diameter. Additionally, a Kevlar ladder structure secures the harness meander and comprises 12 Kevlar runs across $8.4$\,cm, in each package. Refs.\,\onlinecite{2010Cryo...50..465W, 1960RScI...31..660F} and Ref.\,\onlinecite{2009Cryo...49..376V} provide the basis for an estimate of loading in the readout harness and the Kevlar suspension, respectively. The PIPER detector readout harness is NbTi superconducting with copper-nickel cladding in a bundle of twisted pairs. (A related reference\cite{2010Cryo...50..465W} considers a phosphor-bronze clad NbTi harness which is woven into a ribbon.)

The model underestimates the measured loading, but we note that thermal conduction varies across material preparations and manufacturers, and in literature. In Ref.\,\onlinecite{2010Cryo...50..465W}, the phosphor bronze cladding dominates conduction. Applying their range of models to our harness results in loading $0.9-2.0\,{\mu}{\rm W}$, across the range of literature values for conduction of phosphor bronze\cite{2010Cryo...50..465W}. When S2 returns to $T_{\rm recycle}^{\rm S2} = 0.40$\,K to recycle against S3 over $6$\,minutes, the S1 loading increases to $13\,{\mu}{\rm W}$, implying $36\,{\mu}{\rm W}/K$ parasitic conduction across the SCHS.

\subsection{Cycle thermodynamics\label{ssec:thermodynamics}}

% 412 mA on S2, but 362 mA available
The applied coil current is a convenient variable for understanding the thermodynamics of the continuous cycle. When the magnetic field in the crystal exceeds the internal crystal field, 1) the rate of change in entropy, and so power exchanged at constant temperature, is proportional to the change in current, 2) for adiabatic demagnetization of a decoupled stage, the crystal dominates the heat capacity and its temperature directly scales with the magnetic field or, equivalently, coil current. Additionally, the stage controllers directly measure the coil current, so it provides a conveniently measurable quantity that does not require a model for the entropy of the spins or the distribution of the magnetic field in the crystal. Table\,\ref{tab:cadrparams} gives $\Delta T/ \Delta I$ in adiabatic demagnetization, the ratio of current exchanged between CADR stages, stage loading, and the maximum current for each stage.

% previously found 3.5 as charge ratio of S2 to S1
In typical flight operation with a $1.7$\,K CADR baseplate and $100$\,mK detectors, the S1 coil ramp rate is measured to be $5.8$\,mA/min. Hence, over the approximately 30 minute period where the upper stages recycle, $I_{\rm S1}$ falls by $174$\,mA. S1 can be magnetized by at most 800\,mA, so S1 can provide a cooling power several times higher (${\approx}10\,{\mu}{\rm W}$) than needed in PIPER's flight. Additionally, S2 receives approximately $\Delta I_{\rm S2} = +176$\,mA for each recycle with S3, allowing $\Delta I_{\rm S1} = +756$\,mA in exchange with S1 (Table\,\ref{tab:cadrparams}), yielding ${\approx}10\,{\mu}{\rm W}$ cooling in steady-state. Note that higher capacity is possible\cite{2004Cryo...44..581S} through cycle optimization, but is not needed here.

Coil currents can also be used to model the operating temperature range. Starting from S2's exchange with S3 at $400$\,mK and S2 maximum current of $2.5$\,A, S2 can drop to a temperature of $40$\,mK with $100$\,mA remaining. From there, $\Delta I_{\rm S2}=40$\,mA can recharge S1 for continuous operation at $50$\,mK and yield $I_{\rm S2} = 60$\,mA surplus. (In this low-current regime, the self-field effects become more significant, and the cooling power per unit current drops.) In practice, several ADR cycles are required to build the current in S2 necessary to achieve this low-temperature operation, and CADR operation cannot begin until ascent. For each ADR cycle with $\Delta I_{\rm S2} = +176$\,mA, approximately $40$\,mA is needed to re-magnetize S1. The remaining current can be used to drop the temperature of S2 and S1 by roughly $15$\,mK per cycle. Hence after two cycles starting at $100$\,mK, the arrays can operate at $70$\,mK, which we take as a practical lower limit for operation in a $12$\,hr science flight.

Operation of the detector array above $130$\,mK requires an additional step of dropping S1 back to $<130$\,mK before coupling to S2 because the S3-S2 passive switch will become conducting for $T_{\rm S2} > 130$\,mK. S2 can provide an intercept at $120$\,mK ($10$\,mK below S1) without driving the S3-S2 switch to conduction. Continuous operation is therefore limited to $<130$\,mK. These thermodynamic constraints combine to give an operating range $70-130$\,mK with considerable margin in cooling power at $100$\,mK in flight conditions. 

%Behavior in different bath conditions
%Loading compared to estimates, S(T,B) analysis 
%Calibration of hall probes; make sense in terms of field model
%Simulating flight conditions; plot of CADR base temp during flight

\subsection{Stability\label{ssec:stability}}

% from plots guess 1.5 pW/K
%The S1 PID and readout achieve controlled stability of $1.2\,{\mu \rm K}/{\sqrt{\rm Hz}}$, measured at the detector carrier wafer. 
% The detector package thermometry have $3.125\times$ lower excitation ($32$\,nA) than the ADR S1 sensor to avoid temperature offsets from self-heating. 
At the three thermometers on the detector packages, we find $17\,{\mu \rm K}/{\sqrt{\rm Hz}}$ noise. The cross-power spectrum between the thermometers measures coherent temperature fluctuations and is white with level $4.8\,{\mu \rm K}/{\sqrt{\rm Hz}}$. This is consistent with expected noise in the S1 PID control loop, which is based on a thermometry readout which has noise at the level $5.4\,{\mu \rm K}/{\sqrt{\rm Hz}}$.
%The PID loop for S1 uses a thermometer directly on the ADR pill and has readout noise . 

The thermometry is read at $1$\,Hz, so we are unable to directly measure temperature fluctuations in the PIPER science band ($4$\,Hz and harmonics of the VPM modulator). We find that the response of the detector package temperature to fluctuations on the PID current controller is fit with a low-pass filter with one pole at $0.8 \pm 0.2$\,Hz. Hence, $4.8\,{\mu \rm K}/{\sqrt{\rm Hz}}$ fluctuations would roll off to $<0.16\,{\mu \rm K}/{\sqrt{\rm Hz}}$ at frequencies $>4$\,Hz.

Through conduction in the suspensions, the $100$\,mK temperature is sensitive to variations in the baseplate temperature and S4, S3, and S2 pill temperatures. The ratio between variations in these stage temperatures and the $100$\,mK stage temperature (measured at the detector carrier board) is $\{1 \times 10^{-4}, 6 \times 10^{-7}, 6 \times 10^{-6}, 2 \times 10^{-5}\}$, respectively for the baseplate, S4, S3, and S2. The largest is the susceptibility to the baseplate temperature because every suspension connects the ADR stage to the baseplate. Temperature variations in these respective stages during CADR cycling at balloon float are $\{ 0.04, 1.2, 0.9, 0.3 \}$\,K, resulting in temperature variations of $\{ 4.0, 0.7, 5.4, 6.0 \}\,{\mu}{\rm K}$. These deviations are below the level of readout noise, and we determine these susceptibilities using cross-correlation over $20$\,min of data. Deviations follow known templates from the measured stage temperatures, aiding in modeling (as necessary of the TES data). 

For TESs stuck in the superconducting branch after an IV curve, we apply a bias to drive them normal and back onto transition. Maximum biasing heats the detector carrier wafers by up to $70$\,mK, heats the packages by 0.5\,mK and produces a $0.1$\,mK ripple at the PID-controlled S1. The system settles within $5$\,sec, and this represents negligible loading on S1 during regular operation.

% also see:
% http://inspirehep.net/record/1622325/files/Pappas_princeton_0181D_11617.pdf
We can convert temperature fluctuations to equivalent power fluctuations\cite{2018SPIE10708E..3ZC} on the detectors using the susceptibility $G(T_b / T_c)^{n-1}$, where $G$ is the thermal conductivity, $T_b$ and $T_c$ are the bath and critical temperature of the detectors, respectively, and $n$ is the index of the TES conduction law. This yields ${\approx}4.2$\,pW/K to equivalent Joule power so that temperature noise at $<0.16\,{\mu \rm K}/{\sqrt{\rm Hz}}$ becomes $<7 \times 10^{-19}\,{\rm W}/{\sqrt{\rm Hz}}$ equivalent noise in absorbed power. Implied fluctuations are below the dark per-detector median noise-equivalent power (NEP) based on measured TES parameters for array 1 and array 2, $2.4 \times 10^{-18}\,{\rm W}/{\sqrt{\rm Hz}}$ and $4.5 \times 10^{-18}\,{\rm W}/{\sqrt{\rm Hz}}$ respectively, compared to expected photon noise $8 \times 10^{-18}\,{\rm W}/{\sqrt{\rm Hz}}$ (in the $170-230$\,GHz band). $100$\,mK stage temperature fluctuations appear as a common mode across both arrays, aiding in subtraction. Future work aims to improve thermal stability through improved excitation and PID optimization. 
%Cross-correlation of the detector package temperature with the TES time-ordered data isolates true temperature fluctuations from noise in the thermometer and shows susceptibility \red{XYZ}.

\subsection{Detector interaction\label{ssec:interaction}}

% dI/dDAC = 254.54 pA/DAC, dP/dI ~ 35 pW/uA. 
We observe several modes of interaction with the detectors. These appear approximately as a common mode on the detectors, though the susceptibility to radiative, thermal and magnetic effects may vary in detail across the arrays. We will describe the amplitude of the pickup inferred from feedback DAC counts in terms of equivalent (Joule) detector power through a conversion $35\,{\rm pW}/{\mu {\rm A}}$. Most of the effects are also slow compared to the science signal, which is modulated by the $4$\,Hz throw of the VPM.

The S3 magnetic shield in the first flight unit was undersized slightly for the required field range and saturated for fields above $80\%$ of the pill capacity. Due to the difference in temperature between the S4-S3 and S3-S2 passive switches, S3 must be run at its full capacity to recharge S2, leading to a stray field during cyclic operation. Pickup on the detectors produces an equivalent absorbed power ${\approx}1$\,fW that spans ${\approx}3$\,min. In future versions, the shield will be thicker. Additionally, S4 has significant spare capacity, allowing the S4-S3 passive switch temperature to be lowered, reducing the necessary field excursion of S3 during its cycle. The Helmholtz coils of the SCHS are unshielded, but these reach a central field of ${\approx}100$\,mT. There is no evidence of SCHS interference in the time-ordered data.

% power difference in one mode 170-230 GHz from 4K-11K is 10 pW
% in intensity_mapping/EXCLAIM_notebooks/CADR_interaction_analysis.ipynb
% https://arxiv.org/pdf/1808.10491.pdf
% Google TES "bath temperature fluctuations"
% C. Papas thesis
The BP-S4 switch is active and driven by a heated getter on a stem outside of the S4 assembly. We heat the getter to $11$\,K in normal operation to make the switch conducting, and find that this produces a $3$\,fW response in the detectors, implying a coupling of $3 \times 10^{-4}$ into the detector throughput. This detector loading varies slowly over 12\,minutes and is a small fraction of the saturation power ($0.3$\,pW for Arr1 and $1$\,pW for Arr2). Use of a passive heat switch for the 4-BP connection could avoid getter radiation in most systems. However, the PIPER CADR requires an active switch to accommodate changes in bath temperature during the flight. The next iteration of the CADR will directly control radiation from the BP-S4 getter.

%\red{TODO: ask Mark and Peter about this}
% https://journals.aps.org/pr/pdf/10.1103/PhysRev.94.540
S1 alternates between rejecting (during isothermal magnetization against S2) and absorbing heat while its PID control loop seeks to maintain a constant temperature. Cooling from S2 pulls S1 down by $0.24$\,mK during the recharge operation and produces a $1$\,mK transient over ten seconds. The transient effects are rapid and can be masked in ${\approx}20$\,sec of the data across the $30$\,min CADR cycle, representing a small loss of data. In steady-state operation at the base temperature of balloon float, S1 recharge against S2 requires ${\approx}2$\,min of the ${\approx}30$\,min continuous cycle time. Future work will implement a feed-forward controller to respond to heat flow out of S1 from the known temperature difference, and use the PID loop to suppress residual fluctuations. 
% In addition, in its transition from normal metal to superconductor, Pb in the SCHS releases \red{XYZ}\,J, producing a temperature change \red{XYZ} in the 100\,mK stage. 

% 4-BP switch interaction: T4 through back or other through front?
% magnetic field, Model for shield and saturation
% measure TOD x RuOx and scale it to match 1/f features in the RuOx auto-power
% this will be a temperature RMS. Formally this represents an upper bound, because some autopower that looks like coherent structure may not actually correlate with the TOD.

\section{Summary\label{sec:summary}}

We have described the PIPER receiver, which cools two kilo-pixel arrays in a superfluid-tight enclosure in an all-cryogenic telescope in an LHe bucket dewar. This publication is the first report of the performance of a continuous adiabatic refrigerator to cool continuum TES arrays. Control software and electronics are refined to provide highly automatic control of the CADR stages. The CADR provides cooling between $70-130$\,mK with high stability. CADR testing and tuning focused on achieving robust operation and control in flight-like conditions during integrated receiver testing (Sec.\,\ref{ssec:testenv}), and we have outlined areas for future optimization. PIPER is expected to have its first science flight in October 2019 from Ft. Sumner, New Mexico.

%This testing is expensive to carry out for more than one week, and additional performance is likely achievable.

\begin{acknowledgments}
We acknowledge the staff at NASA's Columbia Scientific Balloon Facility for engineering flight operation from Ft. Sumner. We thank Britt Griswold for preparing Fig.\,\ref{fig:piper_complete_receiver} and Nick Bellis for rendering Fig.\,\ref{fig:piper_receiver}. This article may be downloaded for personal use only. Any other use requires prior permission of the author and AIP Publishing. This article appeared in Rev. Sci. Inst. Volume 90(9) and may be found at \verb+https://doi.org/10.1063/1.5108649+.
\end{acknowledgments}

\appendix

%\section{Appendixes}

\nocite{*}
\bibliography{aipsamp}% Produces the bibliography via BibTeX.

\end{document}